\documentclass[a4paper,11pt]{article}
\usepackage{jinstpub}
\usepackage{lineno}
\usepackage{float}
\usepackage{graphicx}
\usepackage{xcolor}
\usepackage{tikz}
\usetikzlibrary{shapes.geometric, arrows.meta, decorations.pathreplacing, arrows, calc, positioning,fit, backgrounds}
\usepackage{mdframed}
\usepackage[separate-uncertainty = true,multi-part-units=single]{siunitx}
\usepackage{needspace}

\DeclareSIUnit\bar{bar}

\title{\boldmath Rasnik 3-point alignment system: algorithm, control framework, and its applications}

\author[a,1]{A. N. Koushik,\note{Corresponding author.}}
\author[b,2]{H. van der Graaf}
\author[d]{K. Ravensberg}
\author[d]{R. M. Wanders} %\orcid{0000-0001-9515-2258}
\author[d]{S. N. Gomashie}
\author[a,2]{N. van Remortel}
\author[b,c]{J. V. van Heijningen}
\affiliation[a]{University of Antwerpen, Groenenborgerlaan 171, 2020 Antwerpen, Belgium}
\affiliation[b]{Nikhef, Science Park 105, 1098 XG, Amsterdam, the Netherlands}
\affiliation[c]{Netherlands Institute for Space Research SRON, Niels Bohrweg 4, 2333 CA, Leiden, the Netherlands}
\affiliation[d]{Department of Physics and Astronomy, Vrije Universiteit, De Boelelaan 1100, 1081 LB, Amsterdam, the Netherlands}

\emailAdd{AnoopNagesh.Koushik@uantwerpen.be}

\abstract{Rasnik is a three-point optical displacement sensor originally developed for particle detector alignment in high-energy physics experiments, including the muon chambers of L3 at LEP and ATLAS at the LHC. The system has evolved from four-quadrant photodiodes to CMOS pixel sensors with custom ChessField coded masks, enabling absolute position measurement with no cumulative drift due to absolute value coded. Key advantages include electromagnetic immunity through purely optical measurement principles, working distances from \SI{50}{\milli\meter} to \SI{15}{\meter}, and multi-degree-of-freedom sensitivity perpendicular to the optical axis. RasCal, a comprehensive control and analysis software, is presented in this paper and its real-time image processing shows \SI{5}{\pico\meter\per\sqrt\hertz} spatial resolution. With GPU acceleration, \SI{274.5}{\hertz} is achieved during live camera acquisition and \SI{109}{\hertz} on CPU. In maximum-throughput configurations, the processing rates exceed \SI{300}{\hertz} on simple consumer hardware. 
System performance is demonstrated across diverse applications: \SI{5}{\pico\meter\per\sqrt{\hertz}} displacement sensitivity is achieved in the VATIGrav setup, dynamic behavior is characterized with a Watt's linkage. In addition, the lack of cumulative drift due to absolute coding with minimal thermal sensitivity under controlled conditions is used for vibration and thermal characterization of photodiode mounts for the LISA space mission. Millisecond-level command latency and thread-safe multi-camera support are provided by the RasCal software, establishing it as a robust and cost-effective precision measurement solution for demanding alignment applications in gravitational-wave detectors and space instrumentation.}

\begin{document}

    \maketitle
    \flushbottom
    
    \section{Introduction}
\label{sec:intro}
    High-precision position sensors are critical components in physics ranging from gravitational-wave detection and particle physics experiments to space applications. The most precise position sensors are interferometric and developed in the gravitational-wave community. A long-range homodyne quadrature interferometer (HoQI)\,\cite{HoQI2016,HoQI2025} has demonstrated displacement Amplitude Spectral Sensity (ASD) of \SI{1e-12}{\meter\per\sqrt{\hertz}} and a few \SI{1e-14}{\meter\per\sqrt{\hertz}} at \SI{10}{\milli\hertz} and \SI{10}{\hertz} respectively, with a dynamic range of several millimeters. Even more sensitive but operating in closed loop is a homodyne interferometer\cite{Gray1999} and further developed as inertial sensor readout at Nikhef, Amsterdam. Attached to a Watt's linkage, \SI{8e-15}{\meter\per\sqrt{\hertz}} inertial displacement sensitivity was obtained from \SI{30}{\hertz} onward\,\cite{Heijningen2018}. The interferometric readout itself showed a reasonably flat \SI{4e-15}{\meter\per\sqrt{\hertz}}\,\cite{Heijningen_PhD}. Longer baseline interferometry, for instance, to monitor the distance between actively isolated optical tables in LIGO, has been developed as suspension point interferometer (SPI)\,\cite{Koehlenbeck2023}, which demonstrated \SI{1e-10}{\meter\per\sqrt{\hertz}} sensitivity over \SI{10}{\meter} at the \SI{10}{\meter} prototype at the Albert Einstein Institute at Hannover; such sensors are now being installed between tables at LIGO\,\cite{LIGOstandard}. 

    However, interferometric sensors only probe one Degree of Freedom (DoF) per optical sensor, which can bring about cross-coupling when combining several devices. In addition, the sensed DoF is always the direction of the optical axis. Examples of sensors that probe the DoFs perpendicular to the optical axis (or axis between sensor elements) are position sensing devices\,\cite{Wallmark1957} and optical levers\,\cite{Jones1961}. While the former sensor class has a typical sensitivity of a fraction of a micron, optical levers can have sub-nanometer resolution. Birmingham Optical Sensor and ElectroMagnetic actuators (BOSEMs), developed specifically for gravitational wave interferometry, achieve displacement noise levels of \SI{4.5e-11}{\meter\per\sqrt{\hertz}} at \SI{1}{\hertz} in their best implementations for the Advanced LIGO A+ upgrade\,\cite{Carbone2012}. However, both optical levers and BOSEMs detect motion of a spot on a photodetector and cannot distinguish between angular motion and translational motion. Both the PSD and optical lever have a dynamic range on the order of the photosensitive area size.

The Rasnik alignment system, originally developed for particle detector alignment in 1983 and successfully deployed in experiments such as L3 at LEP and ATLAS at CERN, addresses many of these fundamental limitations. By encoding absolute position information directly into the optical measurement through a coded mask pattern, Rasnik eliminates the cross-coupling issues of multi-sensor arrangements while maintaining the capability to measure displacements perpendicular to the optical axis, a crucial advantage over purely interferometric approaches.

The 3-point optical alignment system called Rasnik aims to combine sub-nanometer precision with multi-DoF readout with no cross-coupling, including DoFs perpendicular to the optical axis. In addition, the coded mask provides position information, also after shut down and restart of the system. Rasnik achieves sub-nanometer spatial resolution by using a two-dimensional FFT of the image of a custom-made coded mask. This mask is back-lit and imaged onto a pixel sensor by a lens. Position information of one of these Rasnik elements is determined from motion of the projected mask image on the pixel sensor.

RasCal advances Rasnik analysis capabilities through several key innovations beyond previous implementations. During live camera acquisition on the Appendix~\ref{app:pc_specs} hardware, the software achieves near real-time processing at up to \SI{274.5}{\hertz} with GPU acceleration and approximately \SI{109}{\hertz} on CPU alone; in maximum-throughput configurations the GPU processing rate exceeds \SI{300}{\hertz}. The thread-safe multi-camera architecture supports concurrent operation of multiple sensors with independent acquisition and analysis pipelines, essential for large-scale detector alignment monitoring. Integrated live visualization provides immediate feedback through real-time position tracking, frequency-domain noise analysis, and drift monitoring without post-processing delays. The automated data management pipeline handles continuous acquisition with automatic file rotation, data compression, and PDF report generation including comprehensive error analysis and long-term stability metrics. These software innovations, combined with the improved spectral analysis method achieving \SI{5}{\pico\meter\per\sqrt{\hertz}} sensitivity in the VATIGrav setup, establish RasCal as a comprehensive solution for demanding metrology applications.

    Section~\ref{sec:rasnik_system} describes the Rasnik system architecture and coded mask design. Section~\ref{sec:image_analysis} presents the spectral analysis method for position extraction. Section~\ref{sec:linearity_error} presents our findings on velocity-dependent errors. Sections~\ref{sec:control_interface} and~\ref{sec:data_pipeline} detail the software implementation. We conclude with future work on applications and developments.

\section{The Rasnik System}
\label{sec:rasnik_system}
\begin{figure}[h]
        \centering
        \includegraphics[width=0.8\textwidth]{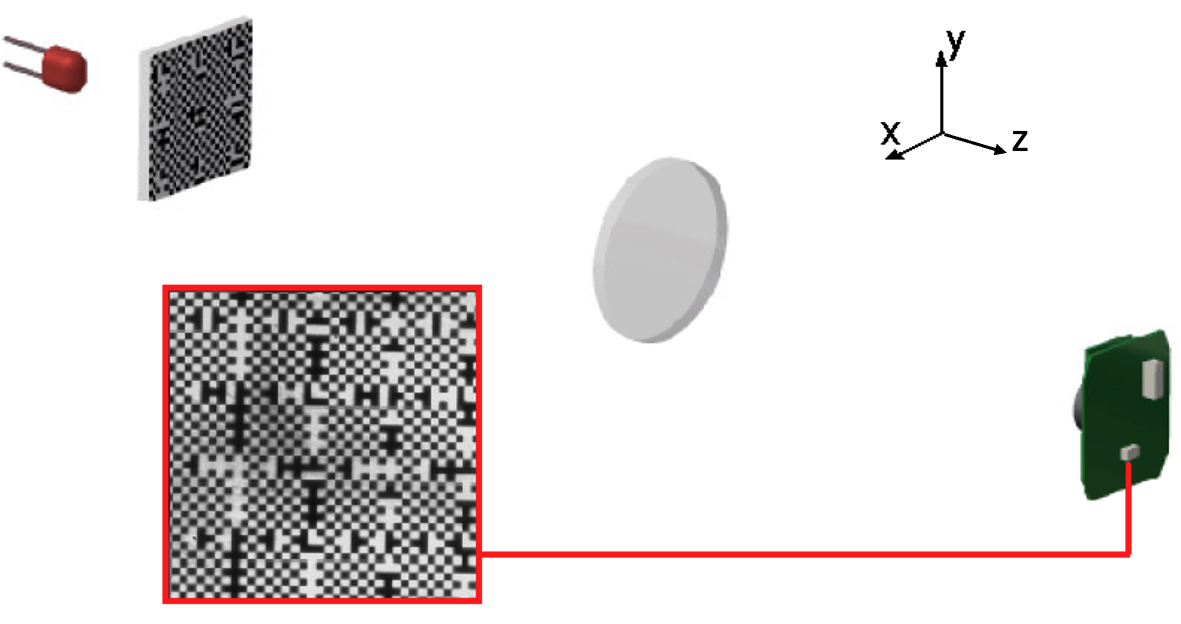}
        \caption{Rasnik: an LED back-illuminates a coded mask and this object is projected onto a pixel sensor using a lens. If any of these three elements move in four degrees of freedom ($x$, $y$, $z$ and roll) the image shifts or rotates and this can be determined using image processing software. The roll (rotation around the $z$-axis) of the lens is not determined. The span of the systems have varied between \SI{50}{\milli\meter} and \SI{15}{\meter}. Adapted from ref.\,\cite{3point}.}
        \label{fig:rasnik_setup}
    \end{figure}

In this section, we describe the Rasnik system architecture, including the optical configuration and coded mask design. The Rasnik system consists of three main components: a back-illuminated coded mask, a lens to project the mask image, and a pixel sensor to capture the image. The system is designed to measure displacements in four degrees of freedom: two transverse translations ($x$ and $y$), one longitudinal translation ($z$), and one rotation (roll) around the optical axis. The roll (rotation around the $z$-axis) of the lens is not determined. The span of the systems have varied between \SI{50}{\milli\meter} and \SI{15}{\meter}. A schematic of the Rasnik setup is shown in Figure~\ref{fig:rasnik_setup}.
\subsection{Optical Configuration}

Since the region of the mask imaged onto the pixel sensor contains position information, it is crucial that the mask be uniformly illuminated. A Lambertian light source would be the optimal choice as it provides uniform angular intensity distribution following Lambert's cosine law, where the radiant intensity varies as the cosine of the angle from the surface normal~\cite{Keiser2022}. This ensures uniform illumination across the entire mask region captured by the pixel sensor, which is crucial for accurate position decoding as the Fourier analysis assumes consistent contrast across the chessboard pattern. The Lambertian characteristic guarantees that the apparent brightness remains constant regardless of viewing angle, eliminating intensity variations that could introduce systematic errors in the frequency spectral analysis. Since only a region of the mask is projected onto the pixel sensors we can use an LED where the optical path starts from the center of the LED. The mask that the LED is illuminating and the pixel sensor capturing is a custom-made mask where the position information within the mask is encoded. This mask is referred to as ChessField mask.

A lens is used to focus the image onto the pixel sensor. The choice of lens depends on the physical setup of the Rasnik readout.  The Rasnik pixel sensor is generally a CMOS sensor. This device is used to capture the image that forms on the CMOS sensor. Typically for Rasnik, a global shutter sensor is used. This reduces image aberrations and artifacts caused by fast movement of the object, which can induce noise in the frequency spectrum, as discussed later in this section. The acquired images are then processed by RasCal (RASnik Control and anALysis), which is standalone DAQ software or can be integrated into existing DAQ systems.

\subsection{Coded Mask Design}

   The mask is similar to a chess board where the ninth row and ninth column are encoded with 8-bit binary value which corresponds to the coarse position on the mask.
    This implies that there are $8 \times 8$ chessboards between every coded values.
    The coarse position of the mask generally starts from the left bottom corner, the origin. From the origin, the ninth row contains the first binary value in x direction. Similarly, the ninth column contains the binary value in y direction. Since all these binary values are 8-bit, there exists one more square, the crossover between the ninth row and the ninth column values which cannot contain any useful coarse position information. This acts as a pivot point by alternating over every set. This block still contributes to the fine positioning calculations.
    
    \begin{figure}[h]
        \centering
        \includegraphics[width=0.5\textwidth]{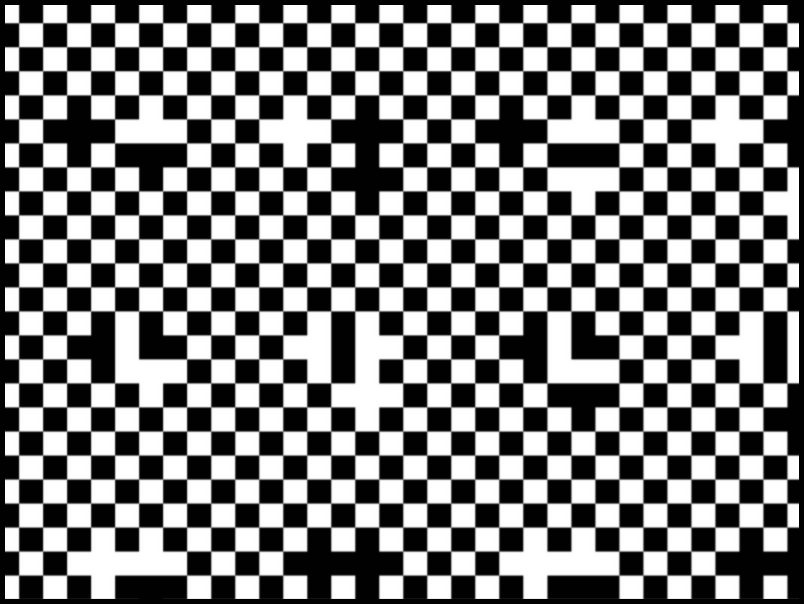}
        \caption{A Rasnik image from a coded mask. The ninth row and ninth column of each $9 \times 9$ block are encoded with 8-bit binary value which corresponds to the coarse position on the mask. 6th row and 7th column in this image counting from top left corner's white square as 1.}
        \label{fig:rasnik_mask}
    \end{figure}
    
    Since 8-bit coded values are used, there is a limit on the size of the mask. The size of the mask can be scaled up or down. This scaling value, the ChessField size, is a necessary parameter in calculating the fine position.
    \\
    The Rasnik setup also uses a lens, which can result in magnification.
    For a well-designed system that produces sharp images, the nominal magnification of a particular setup depends on object and image distance and these are therefore parameters in the image analysis. This allows for the z-coordinate, i.e. displacement along the optical axis, to be an output of image analysis.
    \\
    The final set of parameters needed for the image analysis are:
    \begin{itemize}
        \item $ncode$ (= 9 for the mask used here)
        \item The ChessField size (\SI{4}{\micro\meter} in this work)
        \item The object distance $u$ (distance from lens to mask)
        \item The image distance $v$ (distance from lens to sensor)
        \item The pixel size of the sensor (\SI{3.75}{\micro\meter} for the sensors used in this paper)
        \item The wavelength of the light source ($\lambda = \SI{445}{\nano\meter}$ in this work)
    \end{itemize}

    Masks with different bit values (e.g., 16-bit) are possible with corresponding changes to $ncode$. In this case, the mask would have a larger $ncode$ value, where $ncode = (\text{number of bits}) + 1$. This allows for larger mask sizes but requires adjusting the analysis parameters accordingly.
    This can be accounted in the image analysis routine. Since all the results obtained in this paper are based on an 8-bit mask, the value of $ncode$ remains 9.
    
    The fundamental precision of the Rasnik system is limited by photon shot noise, with the Cramér-Rao lower bound establishing the theoretical limit based on pixel noise and image gradient magnitude. Previous implementations achieved \SI{7}{\pico\meter\per\sqrt{\hertz}}~\cite{ultimate}, while the current work demonstrates improved performance of \SI{5}{\pico\meter\per\sqrt{\hertz}} in the VATIGrav setup, confirming that the system approaches the fundamental limits imposed by the quantum nature of light.

    \section{Image analysis}
\label{sec:image_analysis}
This section describes the pipeline that prepares the image (cropping, binning, windowing) and then estimates absolute position from the ChessField pattern via 2D Fourier analysis and code-square decoding.

\subsection{Analysis Parameters}
\label{sec:analysis_parameters}
The image analysis pipeline requires specific target parameters to ensure compatibility with the spectral analysis algorithms. Parameters such as bit depth and color format can be adjusted based on the specific hardware setup. Spatial resolution plays a crucial role in determining the scale of the ChessField squares in the image; if the input image exceeds the target resolution, additional binning is applied during preprocessing.

For the results presented in this paper, the analysis parameters were set to $720 \times 540$ pixels resolution, 8-bit depth, and monochrome format.

\subsection{Image Preparation}
\label{sec:image_preparation}
The objective of image preparation is to preprocess the acquired images to meet the target specifications defined above while improving data quality and computational efficiency. This involves isolating the Region Of Interest (ROI) from the acquired image, reducing noise, and minimizing spectral artifacts.
The ROI is initially determined either through prior knowledge of the ChessField pattern location or through manual selection.
The following three preprocessing steps are then applied sequentially to achieve optimal spectral analysis:

\begin{enumerate}
    \item \textbf{Cropping:} The ROI is manually configured to ensure the ChessField mask is correctly and completely visible within the analysis window, extracting only the relevant pattern from the full image while discarding extraneous features.
    This careful selection prevents the introduction of artificial phase shifts that could arise from truncating the periodic pattern at arbitrary boundaries.
    Since identical cropping parameters are applied consistently across all images in a measurement sequence, relative position changes between frames are preserved with full accuracy, even though the absolute reference frame is shifted by the cropping operation.
    \item \textbf{Binning:} If the cropped image resolution exceeds the target parameters, adjacent pixels are aggregated into super-pixels, effectively increasing the photon collection per effective sampling element.
    This process reduces photon shot noise and mitigates pixel-to-pixel variations by averaging out sensor inconsistencies.
    Spatial averaging improves the signal-to-noise ratio (SNR) while also decreasing computational load by lowering the effective image resolution.
    \item \textbf{Windowing:} A Hann window~\cite{Harris1978} (with taper parameter tuned to the implementation) is applied to the final ROI, tapering the image edges in the spatial domain.
    This crucial step reduces boundary discontinuities introduced by the cropping operation, thereby minimizing spectral leakage and ringing artifacts in the subsequent Fourier analysis.
\end{enumerate}

\subsection{Spectral Analysis}
\label{sec:spectral_analysis}
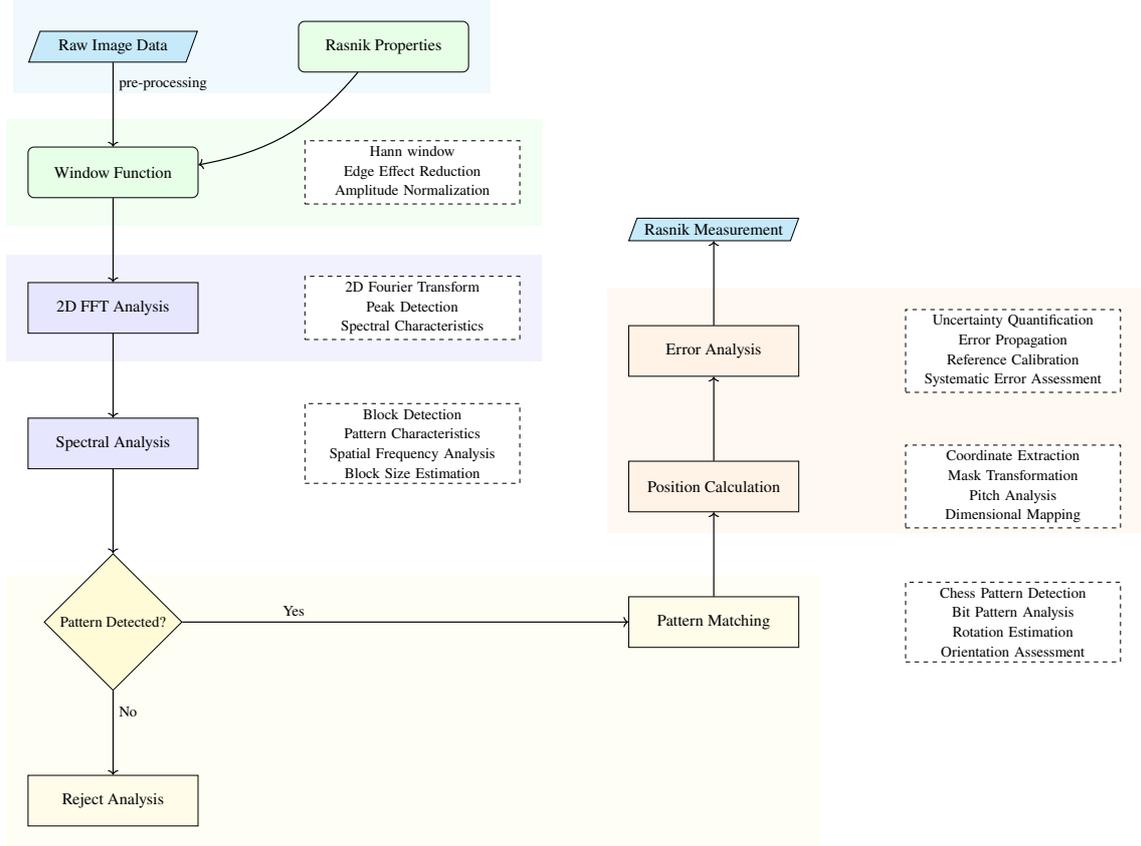
\begin{figure}[htbp]
\centering
    \resizebox{!}{0.5\textheight}{%
    \begin{tikzpicture}[
    auto,
    node distance=2cm and 2.5cm,
    main/.style={rectangle,draw,fill=white,minimum width=4cm,minimum height=1.2cm,thick,text badly centered},
    input/.style={trapezium,trapezium left angle=70,trapezium right angle=110,draw,fill=cyan!20,minimum width=4cm,text badly centered},
    preproc/.style={rectangle,draw,fill=green!10,rounded corners,minimum width=4cm,minimum height=1.2cm,text badly centered},
    spectrum/.style={rectangle,draw,fill=blue!10,minimum width=4cm,minimum height=1.2cm,text badly centered},
    pattern/.style={rectangle,draw,fill=yellow!10,minimum width=4cm,minimum height=1.2cm,text badly centered},
    measure/.style={rectangle,draw,fill=orange!10,minimum width=4cm,minimum height=1.2cm,text badly centered},
    detail/.style={rectangle,draw,dashed,fill=white,text width=4.8cm,align=center,font=\small},
    decision/.style={diamond,draw,fill=yellow!20,minimum width=3cm,minimum height=2cm,font=\small,text badly centered},
    arrow/.style={->,thick}
]

% Column 1: Input and Preprocessing
\node[input] (rawimage) {Raw Image Data};
\node[preproc] (properties) [right=of rawimage] {Rasnik Properties};
\node[preproc] (window) [below=of rawimage] {Window Function};
\node[detail] (windowdetail) [right=of window] {
    Hann window\\
    Edge Effect Reduction\\
    Amplitude Normalization
};

% Column 2: Spectrum and FFT
\node[spectrum] (fft) [below=of window] {2D FFT Analysis};
\node[detail] (spectrum) [right=of fft] {
    2D Fourier Transform\\
    Peak Detection\\
    Spectral Characteristics
};

\node[spectrum] (foam) [below=of fft] {Spectral Analysis};
\node[detail] (foamdetail) [right=of foam] {
    Block Detection\\
    Pattern Characteristics\\
    Spatial Frequency Analysis\\
    Block Size Estimation
};

% Column 3: Pattern and Decision
\node[decision] (pattern) [below=of foam] {Pattern Detected?};
\node[pattern] (search) [below=of pattern] {Reject Analysis};
\node[pattern] (match) [right=of pattern, xshift=8cm] {Pattern Matching};
\node[detail] (matchdetail) [right=of match] {
    Chess Pattern Detection\\
    Bit Pattern Analysis\\
    Rotation Estimation\\
    Orientation Assessment
};

% Column 4: Measurement and Analysis
\node[measure] (position) [above=of match] {Position Calculation};
\node[detail] (posdetail) [right=of position] {
    Coordinate Extraction\\
    Mask Transformation\\
    Pitch Analysis\\
    Dimensional Mapping
};

\node[measure] (error) [above=of position] {Error Analysis};
\node[detail] (errordetail) [right=of error] {
    Uncertainty Quantification\\
    Error Propagation\\
    Reference Calibration\\
    Systematic Error Assessment
};

% Output
\node[input] (result) [above=of error] {Rasnik Measurement};

% Connections
\draw[arrow] (rawimage) -- node[near start, font=\small] {pre-processing} (window);
\draw[arrow] (properties) to[bend left=20] (window);
\draw[arrow] (window) -- (fft);
\draw[arrow] (fft) -- (foam);
\draw[arrow] (foam) -- (pattern);
\draw[arrow] (pattern) -- node[near start, font=\small] {No} (search);
\draw[arrow] (pattern) -- node[near start, font=\small] {Yes} (match);
\draw[arrow] (match) -- (position);
\draw[arrow] (position) -- (error);
\draw[arrow] (error) -- (result);

% Background panels
\begin{pgfonlayer}{background}
    \node[rectangle,fill=cyan!5,fit=(rawimage)(properties),inner sep=0.5cm] {};
    \node[rectangle,fill=green!5,fit=(window)(windowdetail),inner sep=0.5cm] {};
    \node[rectangle,fill=blue!5,fit=(fft)(spectrum),inner sep=0.5cm] {};
    \node[rectangle,fill=yellow!5,fit=(search)(match),inner sep=0.5cm] {};
    \node[rectangle,fill=orange!5,fit=(position)(errordetail),inner sep=0.5cm] {};
\end{pgfonlayer}

\end{tikzpicture}%
}
    \caption{Spectral analysis workflow for Rasnik position measurement. The process begins with image preprocessing (cropping, binning, windowing) followed by 2D FFT analysis to extract spatial frequencies. Peak fitting in the frequency domain determines pattern parameters (period, phase, rotation), which are combined with code square identification to calculate absolute position.}
    \label{fig:rasnik-workflow}
\end{figure}

The spectral analysis method employs a two-dimensional Fourier transform to extract both horizontal and vertical spatial frequencies from the ChessField pattern simultaneously. The workflow in figure~\ref{fig:rasnik-workflow} summarizes the steps. This approach builds upon the fundamental principle that sharp edges in images produce characteristic frequency signatures. By capturing all frequency components in a single transform, the method provides a holistic view of the pattern structure. The spectral approach exhibits robust performance even against degraded images (reduced contrast, added noise, and partial occlusions), maintaining position estimates.

Mathematically, the ChessField mask can be represented as two square waves in perpendicular directions, creating a periodic pattern ideal for frequency domain analysis. The detailed spectral model is described in ref.\,\cite{3point}.

\subsubsection{Two-Dimensional Fourier Analysis}
The preprocessed image $I(x,y)$ is transformed to the frequency domain using the two-dimensional Fast Fourier Transform, denoted $\mathcal{F}_2\{\,\cdot\,\}$, yielding the spectrum $S(u,v)=\mathcal{F}_2\{I\}(u,v)$. The computation employs the FFT algorithm~\cite{Cooley1965} for efficiency, with zero-padding configured to increase the sampling density of the frequency domain. The resulting frequency spectrum is complex-valued, with the magnitude representing the strength of each frequency component and the phase encoding spatial shift information. 

For the two-dimensional pattern, peaks occur at: $$P(i,j) = ((2i-1)f_x, (2j-1)f_y)$$ for all integers $i,j$, where $f_x$ and $f_y$ represent the fundamental frequencies in the horizontal and vertical directions, respectively. The square-wave structure of the ChessField pattern produces odd harmonics with amplitudes proportional to $1/(2i-1)$ and $1/(2j-1)$, enabling the multipeak fitting approach described below. 

The peaks in the spectrum correspond to Gaussian profiles with potentially asymmetric widths when the frequencies in the horizontal and vertical directions differ. The spectral peaks exhibit a characteristic cross-shaped ($+$) profile arising from the separable nature of the horizontal and vertical periodicities. These peaks also encode rotational information, with the cross-shape orientation reflecting the pattern's angular alignment.

\begin{figure}[h]
    \centering
    \includegraphics[width=0.5\textwidth]{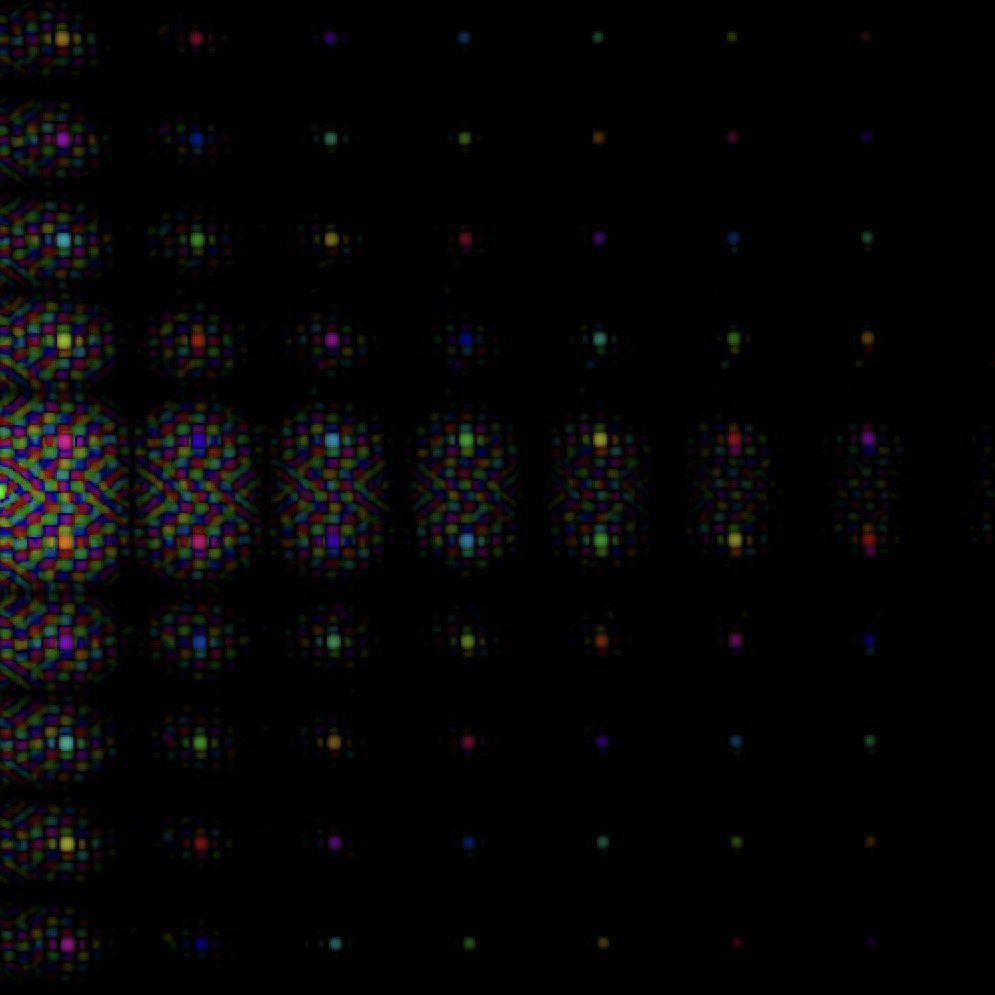}
    \caption{Phase spectrum from the 2D FFT analysis of the ChessField pattern. The zero-shifted representation shows phase values color-coded at each spatial frequency, with peaks at positions $(2i-1)f_x, (2j-1)f_y$ for integers $i,j$. These phase values encode the spatial position of the pattern, enabling sub-pixel measurement precision.}
    \label{fig:fft_image}
\end{figure}

By convention, the zero-frequency component is shifted to the center of the spectrum. For real-valued images, Hermitian symmetry makes half of the 2D spectrum redundant; therefore only the non-redundant half-plane is analyzed. Phase information is typically color-coded for visualization purposes, as illustrated in figure~\ref{fig:fft_image}.

The 2D FFT over an $M \times N$ image scales as $O(MN\log(MN))$~\cite{Cooley1965}.

The resulting spectrum is then used to extract key parameters that are used to decode the position information from the ChessField pattern, as described in the following section.

\subsubsection{Peak Finding and Mask Parameter Extraction}
The frequency domain analysis begins with the identification of significant peaks in the magnitude spectrum $|\mathcal{F}_2(u,v)|$. These peaks correspond to the fundamental spatial frequencies and harmonics of the ChessField pattern. A local maxima detection algorithm identifies peaks within the expected harmonic bands, comparing each candidate to its neighbors, and applying an amplitude cut-off while excluding a small band around the zero axis. Specifically, a point $(u,v)$ is classified as a peak when
\begin{equation}
|\mathcal{F}_2(u,v)| > |\mathcal{F}_2(u+\delta u, v+\delta v)|
\end{equation}
for all $\delta u, \delta v \in \{-1,0,1\}$ excluding $(0,0)$, and additionally satisfies
\begin{equation}
|\mathcal{F}_2(u,v)| > \alpha \cdot \text{median}(|\mathcal{F}_2|),
\end{equation}
where $\alpha$ is a threshold parameter controlling selectivity. This robust peak selection method maintains accurate position determination even under challenging conditions such as partial occlusions or reduced contrast, with the quality metric $S$ (equ.~\ref{eq:skewness}) providing quantitative assessment of pattern integrity.

The identified peaks are sorted by magnitude, with the primary pair taken as the diagonal fundamentals $P(1,1)$ and $P(1,-1)$, which most robustly capture the square-lattice structure in this implementation. These primary peaks should ideally be perpendicular and have equal magnitude. To quantify their geometric relationship, we employ

\begin{equation}
\label{eq:skewness}
S(a,b) = \frac{\lVert a \rVert^2 + \lVert b \rVert^2}{\lvert\det(b,a)\rvert}
\end{equation}

where $a$ and $b$ are the vectors that represent the primary peaks. For perfectly perpendicular peaks of equal magnitude, $S = 2$. Deviations from this ideal value, measured by $\sqrt{S-2}$, serve as a quality metric for the pattern, with values that exceed 0.03 typically indicating significant distortion.

To achieve sub-pixel precision, the integer peak locations are refined using a small-window two-dimensional fit: either a Gaussian model optimized with the Minuit minimization framework \cite{James:1975dr}. These refinements typically yield sub-pixel offsets with uncertainties of 0.001 to 0.01 pixels and reduce the final position uncertainty by approximately an order of magnitude compared to integer-pixel peak detection.

To enhance precision, a multipeak fit is performed that simultaneously considers the primary diagonal peaks $P(1,1)$ and $P(1,-1)$ together with higher harmonics at positions such as $P(3,1)$, $P(1,3)$, and $P(3,3)$. This approach constrains the geometric relationships between peaks, resulting in more accurate parameter estimates. However, the presence of code rows/columns in the ChessField pattern can introduce systematic shifts in peak positions by approximately \SI{1}{\percent} of their width. To account for this uncertainty, an additional error term of 0.0025 frequency bins is incorporated into the peak position estimates\cite{3point}.

From the refined peak parameters, the spatial characteristics of the ChessField pattern are determined, i.e.
\begin{itemize}
    \item The horizontal and vertical spatial frequencies ($f_x$, $f_y$) yield the square sizes in the image
    \item The rotation angle is derived from the orientation of the primary peaks
    \item The phases ($\phi_x$, $\phi_y$) encode the sub-pixel shift of the pattern relative to the image frame
\end{itemize}

The precision of these parameters is quantified from the minimizer’s parameter uncertainties (including correlations handled internally). Typically, the phase can be determined with an uncertainty on the order of 0.001 radians, which translates to a positional precision approximately 3000 times finer than the square size.

\subsection{Absolute Position Calculation}
\label{sec:absolute_position}

The absolute position calculation integrates the phase information from Fourier analysis with the identification of the code squares to establish a coordinate transformation between the image pixels and the physical mask coordinates. This process combines two complementary measurements; sub-pixel shift determination through phase analysis, and coarse positioning via code interpretation from code squares.

The phase values ($\phi_x$, $\phi_y$) extracted from the primary Fourier peaks encode the sub-pixel position of the pattern origin relative to the image coordinate system.
These phases are converted to spatial offsets through

\begin{equation}
\delta_x = \frac{\phi_x T_x}{2\pi}, \quad \delta_y = \frac{\phi_y T_y}{2\pi},
\label{eq:phase_offsets}
\end{equation}

where $T_x$ and $T_y$ represent the spatial periods determined from the Fourier analysis.
These fractional offsets provide nanometer-scale resolution within individual ChessField squares.

Coarse positioning leverages the 8-bit codes encoded in pivot squares located at the intersections of code rows and columns.
The binary pattern formed by these inverted squares encodes integer coordinates $(N_x, N_y)$ that identify which ChessField square contains the image center.
Parity validation ensures robustness against single-bit errors, while spatial consistency checks verify that detected code squares form valid geometric patterns before accepting the decoded position.

The final absolute position combines the coarse integer square counts with the sub-pixel phase offsets, applies the pattern rotation correction, and converts from pixel coordinates to physical units. Residual non-orthogonality is limited during peak selection/fitting via a maximum-skew constraint rather than an explicit post-correction.
\begin{equation}
\begin{bmatrix}
X_{\mathrm{abs}} \\
Y_{\mathrm{abs}}
\end{bmatrix}
= \mathbf{K} \cdot \mathbf{R}(\theta) \cdot
\left(
\begin{bmatrix}
N_x T_x \\
N_y T_y
\end{bmatrix}
+
\begin{bmatrix}
\delta_x \\
\delta_y
\end{bmatrix}
\right),
\label{eq:absolute_position}
\end{equation}
where $(N_x, N_y)$ are the integer square indices from code decoding, and $(T_x, T_y)$ are the spatial periods in pixels from the Fourier analysis. The scale matrix $\mathbf{K}$ converts from pixel to physical coordinates:
\begin{equation}
\mathbf{K} = \begin{bmatrix} k_x & 0 \\ 0 & k_y \end{bmatrix} = \begin{bmatrix} p/m_x & 0 \\ 0 & p/m_y \end{bmatrix},
\end{equation}
where $p$ is the physical mask pitch (square size on the ChessField mask), $k_{x}$ and $k_{y}$ are the pixel-to-physical scaling factors obtained from image sensor pixel size and $(m_x, m_y)$ are the calibrated optical magnifications in the horizontal and vertical directions. The rotation matrix $\mathbf{R}(\theta)$ accounts for pattern orientation determined from the spectral peak analysis:
\begin{equation}
\mathbf{R}(\theta) = \begin{bmatrix} \cos\theta & -\sin\theta \\ \sin\theta & \cos\theta \end{bmatrix},
\end{equation}
Each image typically contains multiple code squares; the analysis selects the best-fitting code value (minimizing inconsistent squares) and applies the phase offsets $(\delta_x, \delta_y)$ to this reference position for the final absolute position calculation.

Uncertainties from each component (FFT period and phase determination, code square localization, and optical magnification calibration) are combined from the respective fit uncertainties to obtain the final position uncertainty.
Experimental validation demonstrates that position determination with relative uncertainties below \SI{0.5}{ppm} is achieved for mask-to-sensor distances between \SI{50}{\milli\meter} and \SI{200}{\milli\meter}~\cite{ultimate}.

\subsection{Computational Performance and Real-Time Viability}
\label{sec:computational_performance}

The spectral analysis method's computational efficiency, combined with modern hardware acceleration, enables real-time position monitoring essential for gravitational wave detector alignment control. The processing time for image analysis directly impacts the system's viability for real-time alignment monitoring. The implementation was benchmarked on a standard workstation (specifications in Appendix~\ref{app:pc_specs}) comparing CPU-based FFT against GPU-accelerated implementations using NVIDIA's cuFFT library~\cite{NVIDIA2023}.

\begin{figure}[h]
    \centering
    \includegraphics[width=\textwidth]{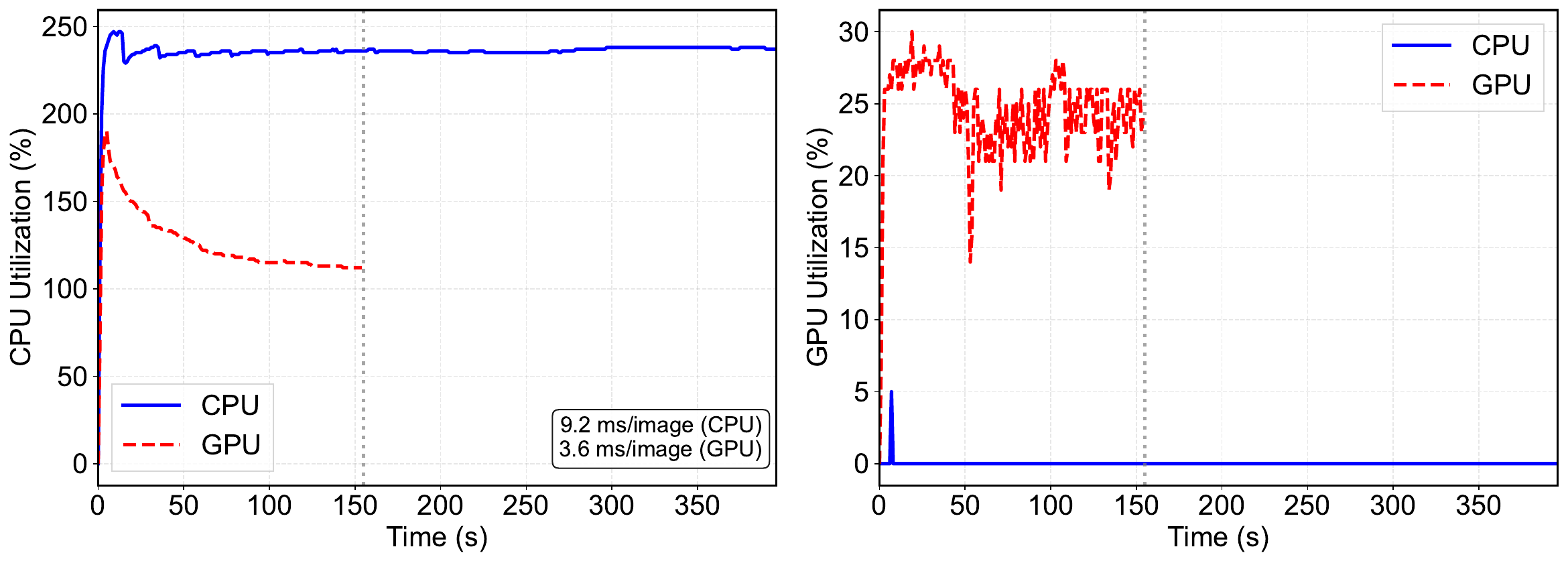}
    \caption{System resource utilization during Rasnik image processing. (a) CPU utilization showing \SI{235}{\percent} usage for CPU-only implementation (blue) versus \SI{115}{\percent} for GPU-accelerated implementation (red). (b) GPU utilization showing \SI{25}{\percent} usage for GPU implementation and zero for CPU-only. The vertical dotted line at 154 seconds marks GPU processing completion. Both implementations processed 43,000 images, achieving \SI{9.2}{\milli\second}/image (CPU) and \SI{3.6}{\milli\second}/image (GPU) respectively.}
    \label{fig:compute_benchmark}
\end{figure}

For typical $720 \times 540$ pixel images, the CPU implementation requires approximately \SI{9.2}{\milli\second} per image, limiting throughput to approximately \SI{109}{\hertz}. The GPU-accelerated version reduces this to \SI{3.6}{\milli\second} per image, enabling processing rates of approximately \SI{278}{\hertz} in queued maximum-throughput tests, well above typical pixel sensor acquisition rates. During live camera acquisition on the Appendix~\ref{app:pc_specs} hardware, the observed maximum processing rate is \SI{274.5}{\hertz} on GPU. figure~\ref{fig:compute_benchmark} demonstrates this performance improvement, showing CPU and GPU usage patterns when processing a queue of 43,000 images. In maximum-throughput configurations, GPU processing exceeds \SI{300}{\hertz}. The 2.6-fold speedup is achieved with minimal code modification.

The CPU implementation's processing rate of \SI{100}{\hertz} exceeds typical pixel sensor frame rates of \SI{30}{\hertz} to \SI{60}{\hertz}, providing adequate performance for single-pixel sensor systems. The GPU-accelerated implementation, with its \SI{3.6}{\milli\second} processing time per image, can theoretically handle frame rates exceeding \SI{275}{\hertz}, ensuring sufficient computational capacity even when processing multiple Rasnik pixel sensors simultaneously. This computational headroom enables real-time position monitoring across distributed sensor networks, as required in large-scale gravitational wave detector alignment systems.

    \subsection{Linearity Error Analysis and Mitigation}
\label{sec:linearity_error}

Monte Carlo simulations reveal that Rasnik's linearity error can reach \SI{40}{\nano\meter}, representing a fundamental limitation 40 times larger than its spatial resolution of approximately \SI{1}{\nano\meter}. This periodic error pattern, first characterized by van der Graaf et al.~\cite{ultimate}, remains the primary constraint on system performance~\cite{3point}.

The Rasnik system exhibits two distinct types of proportionality errors~\cite{3point}. The slope error, a systematic deviation from expected proportionality, is determined by the mechanical precision of the mask and pixel geometry and is expected to be less than \SI{1e-6}{} for masks and sensors with approximately \SI{1}{\milli\meter} sides due to nm-precision MEMS technology. More critically, the position-dependent error shows systematic deviations that depend on the image position on the sensor, which forms the primary concern of this analysis.

The linearity error originates from the discrete phase extraction methodology in the spectral analysis algorithm~\cite{ultimate}. The periodic character of the error strongly suggests that determining peak positions in 2D Fourier plots is a weak point in the current analysis routine. When the ChessField pattern is sampled at discrete pixel locations, the discrete sampling introduces periodic errors in the extracted phase. Since position is calculated from phase via
\begin{equation}
\label{eq:phase_to_position}
x = \frac{\phi \, T_p}{2\pi},
\end{equation}
where $T_p$ is the pattern period, these phase errors directly manifest as position-dependent measurement errors with spatial period matching the ChessField pattern. The error magnitude depends critically on the pattern-pixel alignment: when the ChessField pattern aligns perfectly with the pixel grid (differential rotation $\theta_z = 0$), coherent systematic errors reach their maximum amplitude.

\begin{figure}[htbp]
    \centering
    \includegraphics[width=0.8\textwidth]{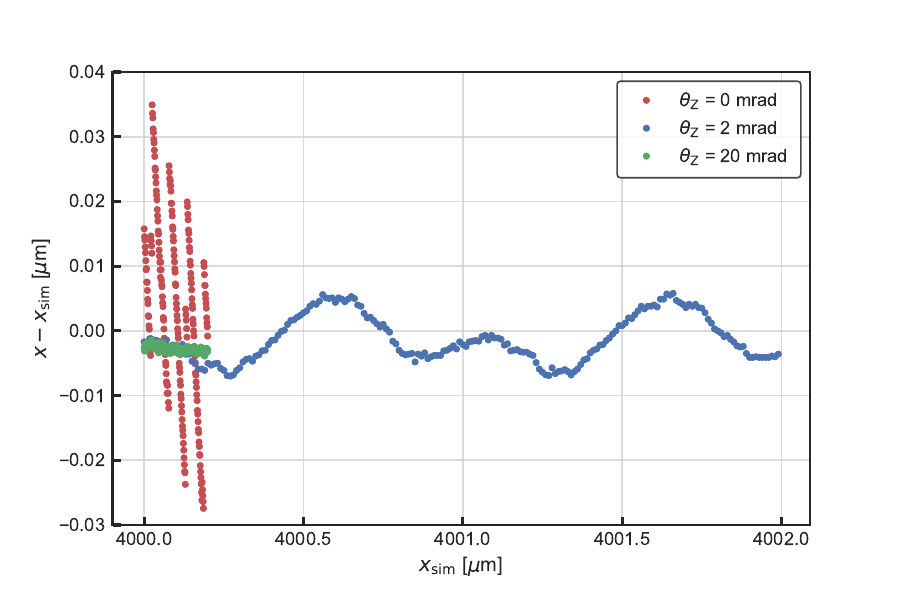}
    \caption{Linearity error as a function of displacement for different rotation angles $\theta_z$. The error exhibits strong periodicity matching the ChessField pattern when $\theta_z = \SI{0}{\milli\radian}$ (maximum amplitude \SI{35}{\nano\meter} to \SI{40}{\nano\meter}), with amplitude decreasing as rotation increases. At $\theta_z = \SI{20}{\milli\radian}$, the error is reduced to minimal levels. Reproduced from~\cite{ultimate}.}
    \label{fig:linearity_error}
\end{figure}

Characterization through both Monte Carlo simulations and experimental measurements demonstrates strong dependencies on the differential angular rotation $\theta_z$ between mask and pixel sensor~\cite{ultimate}. figure~\ref{fig:linearity_error} shows that with $\theta_z = \SI{0}{\milli\radian}$, the maximum error amplitude reaches \SI{35}{\nano\meter} to \SI{40}{\nano\meter} peak-to-peak. At $\theta_z = \SI{2}{\milli\radian}$, the error amplitude reduces to approximately \SI{10}{\nano\meter} peak-to-peak while maintaining a clear periodic pattern. With $\theta_z = \SI{20}{\milli\radian}$, the error amplitude becomes minimal, effectively eliminating the periodic structure.

The dramatic improvement with non-zero $\theta_z$ occurs through frequency domain delocalization. When the ChessField pattern is rotated relative to the pixel grid, phase information becomes distributed across multiple pixels, effectively implementing spatial dithering that decorrelates systematic errors. This averaging effect reduces the error amplitude from \SI{40}{\nano\meter} at zero rotation to approximately \SI{10}{\nano\meter} at \SI{2}{\milli\radian} rotation, and to minimal levels at \SI{20}{\milli\radian}.

The errors originate from the spectral method's discrete phase extraction. When FFT bin quantization constrains peaks to integer locations, it creates staircase-like errors of order $\pi/N$ where $N$ is the FFT size. This effect compounds when the 5-point parabolic interpolation encounters peaks between pixels, where the actual peak shape follows a sinc function, $\text{sinc}(x) = \sin(\pi x)/(\pi x)$, deviating from the quadratic assumption. Phase unwrapping further complicates matters by introducing artificial discontinuities at $\pm\pi$ boundaries when the phase is wrapped to the range $[-\pi, \pi]$, particularly problematic near pattern edges where phase gradients are steep.

\begin{figure}[htbp]
    \centering
    \includegraphics[width=0.8\textwidth]{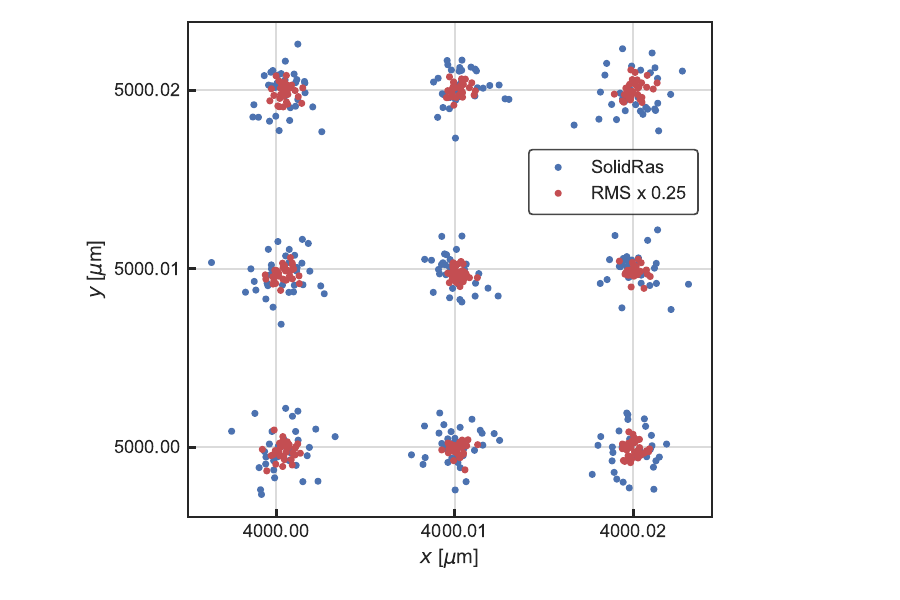}
    \caption{Spatial pattern of linearity error showing 3 $\times$ 3 grid measurements. Non-linear response is visible across the measurement field, with spread at each point due to pixel content fluctuations. Reproduced from~\cite{ultimate}.}
    \label{fig:spatial_pattern}
\end{figure}

The spatial pattern of linearity errors has been well-characterized~\cite{ultimate}. figure~\ref{fig:spatial_pattern} shows the 3 $\times$ 3 grid measurements from~\cite{ultimate} displaying non-linear response across the measurement field, with spread at each point attributed to pixel content fluctuations. The reduced pixel quantum fluctuations are demonstrated by RMS$\times$0.25 scaling.

In practice, linearity errors are minimized through established methods~\cite{3point}. A differential rotation $\theta_z \geq \SI{2}{\milli\radian}$ between mask and sensor reduces errors from \SI{40}{\nano\meter} to approximately \SI{10}{\nano\meter}, achievable through mechanical adjustment during installation or intentional mask mounting with slight rotation. For applications requiring higher precision, post-processing corrections can further reduce residual errors by characterizing and compensating for the periodic error pattern through calibration sweeps.

Future algorithm enhancements must address the fundamental weakness in 2D Fourier peak position determination. Priority improvements include implementing sinc-based interpolation using larger kernels (minimum $7 \times 7$), iterative refinement of peak positions using Newton-Raphson methods, joint estimation of all parameters using maximum likelihood frameworks, and detailed analysis of the 2D Fourier procedures in the spectral analysis algorithm. These improvements could potentially reduce the residual linearity error below \SI{1}{\nano\meter}, approaching the system's inherent spatial resolution limit.

The contribution of linearity errors to the total error budget depends strongly on operating conditions. With proper rotation ($\theta_z \geq \SI{2}{\milli\radian}$) and post-processing corrections, the linearity error contribution can be reduced from a dominant \SI{40}{\nano\meter} to approximately \SI{10}{\nano\meter}, though this may still exceed other error sources such as temperature drift and remains well above the shot-noise-limited performance of \SI{7}{\pico\meter\per\sqrt{\hertz}}~\cite{ultimate}.

\begin{figure}[ht!]
    \centering
    \includegraphics[width=0.8\textwidth]{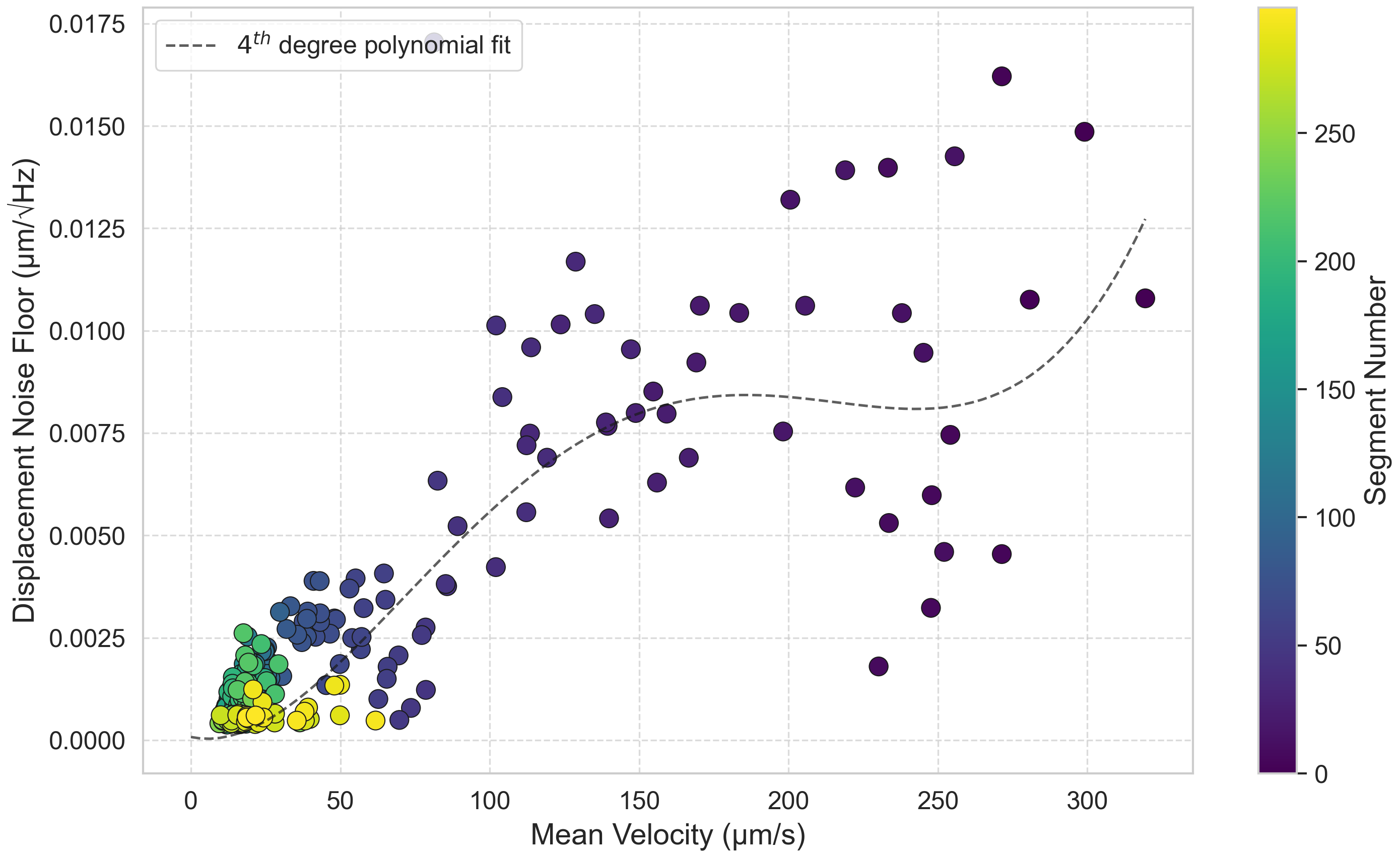}
    \caption{y-axis position data from an oscillating Watt's linkage system showing linearity error manifestation. Where the speed of the object is large, the displacement during two moments of measurement introduces a linearity error that appears as noise. This error resembles the well-known movement error due to dynamic image deformation, but its origin is entirely different, rooted in the discrete phase extraction of the analysis algorithm.}
    \label{fig:velocity_noise}
\end{figure}

The dynamic manifestation of linearity errors was experimentally investigated using a novel Watt's linkage mechanism for controlled periodic motion. This experimental setup provides direct measurement of linearity errors under dynamic conditions, complementing the static Monte Carlo simulations. The experiments reveal that linearity errors become prominently visible as velocity-dependent noise during motion. figure~\ref{fig:velocity_noise} shows that when the object speed is large, the displacement between successive measurements introduces systematic errors that appear as measurement noise. This error resembles movement error from dynamic image deformation but has an entirely different origin rooted in the discrete phase extraction of the analysis algorithm. For the Watt's linkage experiments specifically, magnetic eddy-current damping was employed to constrain motion and reduce oscillation artifacts from the camera, limiting the dynamic range to regions of acceptable linearity.

These dynamic errors can be modeled through a position-dependent transfer function $H(x) = 1 + \varepsilon \cdot \sin(2\pi x/\Lambda)$, where $\varepsilon$ represents a small error amplitude and the error manifestation is proportional to the displacement rate: $\delta x_{error} \approx \varepsilon \cdot \sin(2\pi x/\Lambda) \cdot (dx/dt)$. This finding has important implications for dynamic applications, particularly in gravitational wave detectors where continuous motion can generate position errors comparable to or exceeding the static linearity errors. Addressing both static and dynamic linearity errors is crucial for sub-nanometer precision in next-generation optical metrology systems.

    \section{Control and Interface}
\label{sec:control_interface}

\subsection{Camera Plugin}
The control software allows usage of different types of cameras, such as the Pi camera or USB camera, and supports multiple cameras.
Along with the cameras support, the software is also designed to read the stream of images from the camera over the network or from a file on disk.

The original software includes a camera plugin developed to support the Daheng MER2 series camera, due to its specific requirements such as its high frame rate, resolution, global shutter which is necessary to avoid image distortion and aberrations, and monochrome capability.
The Daheng camera plugin provides complete control of the camera settings such as fps, exposure time, gain, and the pixel format.

The software supports simultaneous operation of multiple cameras limited by USB bandwidth and processing capacity rather than software constraints. Multiple image analysis pipelines can process images from multiple cameras, with live monitoring capabilities for up to 2 cameras.

The software also supports the Pi camera and USB camera, which can be used for testing and development purposes.

\subsection{Image processing delay}
Since using the Rasnik displacement sensor in a closed loop is possible, understanding the delay in obtaining the value is crucial. Unlike the analog sensors where the signal obtained is instantaneous, there is a comparatively significant delay due to multiple steps involved to compute the position value. Two factors that contribute to the delay are discussed in the following.

\paragraph{Data throughput}
Sub-millisecond command latency is achieved through asynchronous processing and memory-mapped I/O. Data transfer delays include constant hardware delays from cable length and protocol overhead from USB bus bandwidth saturation.

\paragraph{Analysis pipeline queue}
The time spent for the image in the queue of the analysis pipeline. This is generally almost instantaneous for analog sensors. However, in digital imaging systems, when multiple frames arrive faster than they can be processed, images accumulate in the queue resulting in increased latency. The implementation prioritizes speed while maintaining accuracy of the analysis pipeline to minimize this delay.

\subsection{Application interface}
The control interface is designed with modern web technologies to provide platform-independent access and control. The front-end utilizes Bootstrap framework for responsive design, ensuring usability across desktop and mobile devices. The web interface provides live visualization of camera feeds, and system status.

A RESTful API (RasAPI) facilitates programmatic access to all system functions, enabling integration with external control systems and automation frameworks. The API follows standard HTTP conventions with JSON data exchange and access control mechanisms. This architecture allows for both human operator control via the web interface and machine-to-machine communication through the API endpoints.

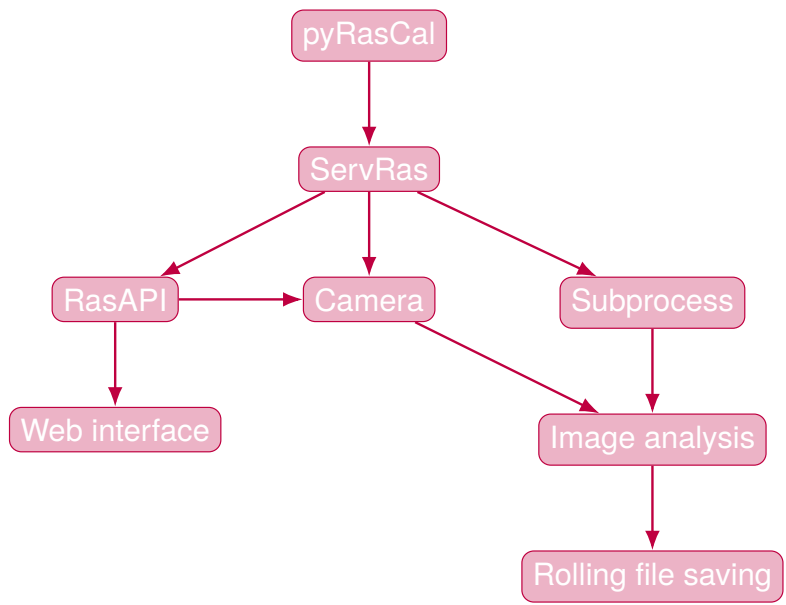
\begin{figure}[h]
    \centering
    \resizebox{!}{0.35\textheight}{%
        \begin{tikzpicture}[
    node distance=1cm and 1.4cm,
    every node/.style={rectangle, rounded corners, draw=purple, fill=purple!30, text centered, text=white, font=\sffamily},
    arrow/.style={-{Latex[scale=1]}, thick, purple}
]

% Nodes
\node (pyRasCal) {pyRasCal};
\node (ServRas) [below=of pyRasCal] {ServRas};
\node (RasAPI) [below left=of ServRas] {RasAPI};
\node (Subprocess) [below right=of ServRas] {Subprocess};
\node (WebInterface) [below=of RasAPI] {Web interface};
\node (Camera) [below=of ServRas] {Camera};
\node (ImageAnalysis) [below=of Subprocess] {Image analysis};
\node (RollingFileSaving) [below=of ImageAnalysis] {Rolling file saving};

% Arrows
\draw [arrow] (pyRasCal) -- (ServRas);
\draw [arrow] (ServRas) -- (RasAPI);
\draw [arrow] (ServRas) -- (Subprocess);
\draw [arrow] (RasAPI) -- (WebInterface);
\draw [arrow] (ServRas) -- (Camera);
\draw [arrow] (Camera) -- (ImageAnalysis);
\draw [arrow] (Subprocess) -- (ImageAnalysis);
\draw [arrow] (ImageAnalysis) -- (RollingFileSaving);
\draw [arrow] (RasAPI) -- (Camera);

\end{tikzpicture}%
    }
    \caption{pyRasCal workflow}
    \label{fig:pyrascal-workflow}
\end{figure}

As illustrated in figure~\ref{fig:pyrascal-workflow}, the interface layer is an integral part of the pyRasCal system architecture. The core ServRas component provides centralized management, bridging the RasAPI, camera subsystem, and image processing pipeline. During live camera acquisition on the Appendix~\ref{app:pc_specs} hardware, GPU-accelerated processing reaches up to \SI{274.5}{\hertz} (CPU \SI{109}{\hertz}); in maximum-throughput configurations the GPU processing rate exceeds \SI{300}{\hertz}. Typical operation maintains memory footprint below \SI{500}{\mega\byte} with \SI{115}{\percent} CPU utilization compared to \SI{235}{\percent} for CPU-only operation. The RasAPI component directly interfaces with both the web interface and camera control systems, allowing for seamless operation through either human or programmatic control.

The web interface provides intuitive access to all system features including:
\begin{itemize}
    \item Live pixel sensor monitoring and configuration
    \item Adjustment of acquisition parameters (exposure, gain, frame rate)
    \item Selection between full image and binned image processing modes
    \item Visualization of analysis results in displacement and ASD plots
    \item Submit background jobs via web interface to generate reports for given time interval and selected options for post processing the data
\end{itemize}

For integration with external data acquisition systems, the RasAPI exposes endpoints for camera control, processing pipeline configuration, and data retrieval. Performance benchmarks detailed in Section~\ref{sec:computational_performance} demonstrate the system's capability for real-time operation. This enables the pyRasCal system to function both as a standalone application and as a component within larger experimental setups.

\subsubsection*{Front-end}
User interface is designed to be simple and intuitive. The interface is divided into four main sections:
\begin{itemize}
    \item \textbf{Camera Control and monitoring:} This page allows control of camera settings such as exposure time, gain, frame rate, and pixel format. The live feed from the camera and image histogram are visible for adjusting the mask position in the setup. As sharp mask images are required for optimal performance, this live feedback facilitates precise mask positioning.
    \begin{figure}[H]
        \centering
        \includegraphics[width=0.8\textwidth]{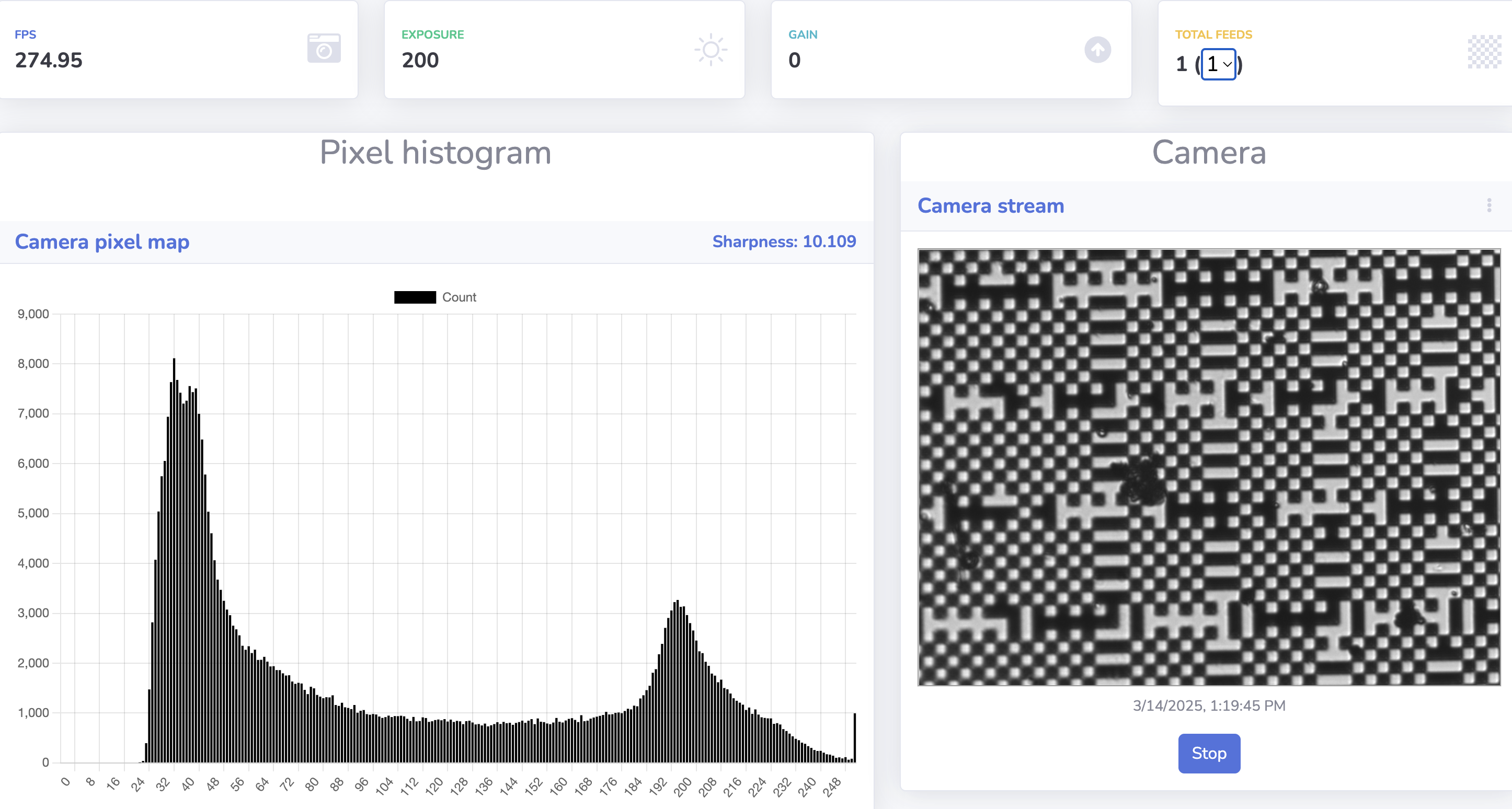}
        \caption{Camera control interface. Left: pixel intensity histogram showing bimodal distribution characteristic of ChessField patterns. Right: live camera feed displaying the Rasnik mask with encoded position information (sharpness: 10.109). Top indicators show current frame rate (\SI{274.95}{\hertz}), exposure (\SI{200}{\micro\second}), and gain (0~dB) settings. A blob of dust is visible in the image which is handled by the image analysis algorithm without impacting position measurement.}
        \label{fig:camera-control}
    \end{figure}
    \item \textbf{x-y displacement:} This page of the software plots the \(x\) and \(y\) position from the image analysis pipeline. The user can view the live position as it is plotted as a timeseries. This page also provides the user to toggle Measurement mode which will continuously process the new images in the background and save the data to disk.
    % \begin{figure}[H]
    %     \centering
    %     \includegraphics[width=0.8\textwidth]{images/rascal/rascal_plotting}
    %     \caption{Displacement plotting page}
    %     \label{fig:displacement-plotting}
    % \end{figure}
    \item \textbf{ASD:} This page of the software plots the amplitude spectral density of the displacement data for both the \(x\) and \(y\) position. The user can view the ASD plot which is updated every 30 seconds.
    % \begin{figure}[H]
    %     \centering
    %     \includegraphics[width=0.8\textwidth]{images/rascal/rascal_asd}
    %     \caption{ASD plotting page}
    %     \label{fig:asd-plotting}
    % \end{figure}
    \item \textbf{Analysis/Reports:} This page of the software allows the user to submit a background job to generate reports for a given time interval. The user can select the options for post-processing the data such as filtering, averaging, and plotting the data. Section~\ref{sec:data_pipeline} discusses the post-processing techniques in more detail.
\end{itemize}

    \section{Data Pipeline}
\label{sec:data_pipeline}
\subsection{Data quality checks and integrity}
Since the data acquisition system is a pixel sensor, the data obtained is not periodic.
The commercial pixel sensors maintain a constant frame rate when averaged over a long period of time.
For general usage of the camera, drops in frame rate do not have a significant impact.
Since here it is used as a sensor, maintaining periodic data is crucial.

\subsubsection{Data quality checks}
\label{sec:data_quality}
The data obtained from the pixel sensor is subjected to various checks to ensure the data quality.
This includes checking for glitches, spikes, null values, and missing data.

\paragraph{Glitches and spikes}
Glitches can arise from transient image corruption (e.g., sudden illumination changes, partial occlusions, or sensor saturation), which can disrupt decoding and yield spurious coarse-code values and extreme position outliers.

In the acquisition pipeline, problematic frames may be flagged or discarded; however, a dedicated post-processing data quality stage enforces robust outlier rejection. Specifically, values are clipped when they exceed a user-selected multiple of the standard deviation (e.g., $k\sigma$) computed over a stable segment of data. This clipping removes extreme outliers while preserving the bulk distribution of valid measurements and is applied before spectral analyses and statistical summaries.

\paragraph{Missing data}
Instances of missing data are handled in two distinct contexts. For visualization, an Exponentially Weighted Moving Average (EWMA) with time constant $\tau=\SI{0.1}{\second}$ provides visual smoothing only and is not used for quantitative spectral estimates:
\begin{equation}
\label{eq:ewma}
y[n] = \alpha \, x[n] + (1-\alpha) \, y[n-1], \quad \alpha = 1 - \exp\!\left(-\frac{\Delta t}{\tau}\right).
\end{equation}
For Amplitude Spectral Density (ASD) estimation, we apply the clipped values (per the $k\sigma$ rule above) and perform a forward-fill substitution only at glitch timestamps with missing samples, i.e., a missing sample is replaced by the last valid measurement until a new valid sample arrives. This preserves sample timing and avoids introducing artificial frequency components beyond those implicit in the hold operation. The distinction between visualization (EWMA) and analysis (clipping and forward fill for missing samples) is maintained explicitly throughout.

\subsubsection{Data storage methodology}
Our data storage strategy involves dividing the continuous stream of sensor data into hourly files using a rolling file mechanism. This approach allows for efficient data management and reduces computational overhead. The system is designed to prioritize computational efficiency by directly writing data to files without intermediate processing. This ensures that the system can handle the high data rate of the acquisition system. This method optimizes both storage usage and minimizes processing delays, which are important for continuous, high-frequency data collection in precision measurement applications.

% \subsection{Linearity error}
% \begin{figure}[H]
%     \centering
%     \includegraphics[width=\textwidth]{images/xy_move}
%     \caption{Linearity error}
%     \label{fig:linearity_error}
% \end{figure}

    \section{Applications}
\label{sec:applications}

The Rasnik system's unique capabilities of sub-nanometer resolution, non-contact operation, dynamic measurement capability, and its sensitivity to the axes perpendicular to the optical axis make it well suited for demanding applications in gravitational wave physics and verification programs for space missions. This perpendicular-axis sensitivity distinguishes Rasnik from interferometric sensors, which only probe along their optical axis, enabling comprehensive multi-DoF monitoring essential for complex alignment tasks. Two key application domains where Rasnik has demonstrated significant impact are presented in this section.

\subsection{Gravitational Wave Detector Instrumentation}
\label{sec:gw_applications}

While Rasnik systems are not currently deployed in operational gravitational wave detectors, prototype testing has been conducted to explore their potential for future detector upgrades. These investigations focus on exploiting Rasnik's unique capability to measure displacements perpendicular to the optical axis, complementing existing interferometric sensors.

\subsubsection{Watt's Linkage Testing for Dynamic Characterization}
\label{sec:watts_linkage}

In what follows, we refer to the Rasnik readout on a Watt's linkage configuration as \emph{RasWatt}. RasWatt performance was evaluated in a Watt's linkage testbed to quantify the intrinsic optical readout sensitivity and the inertial behavior of the mechanical stage. Two complementary measurements are presented: a locked proof-mass configuration to reveal the optical readout noise floor, and an unlocked configuration to show the linkage resonance and inertial response. Measurements were performed at the VATIGrav vacuum facility at the University of Hamburg.

The VATIGrav facility employs a hierarchical two-stage isolation architecture designed to minimize seismic coupling to precision optical measurements. The vacuum chamber (\SI{1.02}{\meter} $\times$ \SI{1.74}{\meter} $\times$ \SI{1.51}{\meter} internal volume) is supported by four STACIS III active vibration isolators that provide broadband suppression above \SI{0.2}{\hertz}, with translational noise below \SI{e-7}{\meter\per\sqrt{\hertz}} achieved above \SI{10}{\hertz}. Inside the chamber, a \SI{240}{\kilo\gram} optical table is supported by four Minus-K 250 CM-1 passive isolators, each weighing \SI{12}{\kilo\gram} and tuned to a resonance frequency near \SI{0.5}{\hertz}. This dual-stage configuration provides horizontal motion attenuation up to two orders of magnitude at frequencies above the passive isolator resonances. Vacuum pressures of \SI{e-1}{\milli\bar} are maintained using oil-free scroll pumps and a magnetically levitated turbomolecular pump, ensuring hydrocarbon-free conditions essential for optical metrology~\cite{Basalaev2025,Gerberding2024}.

\begin{figure}[htbp]
    \centering
    \includegraphics[width=0.85\textwidth]{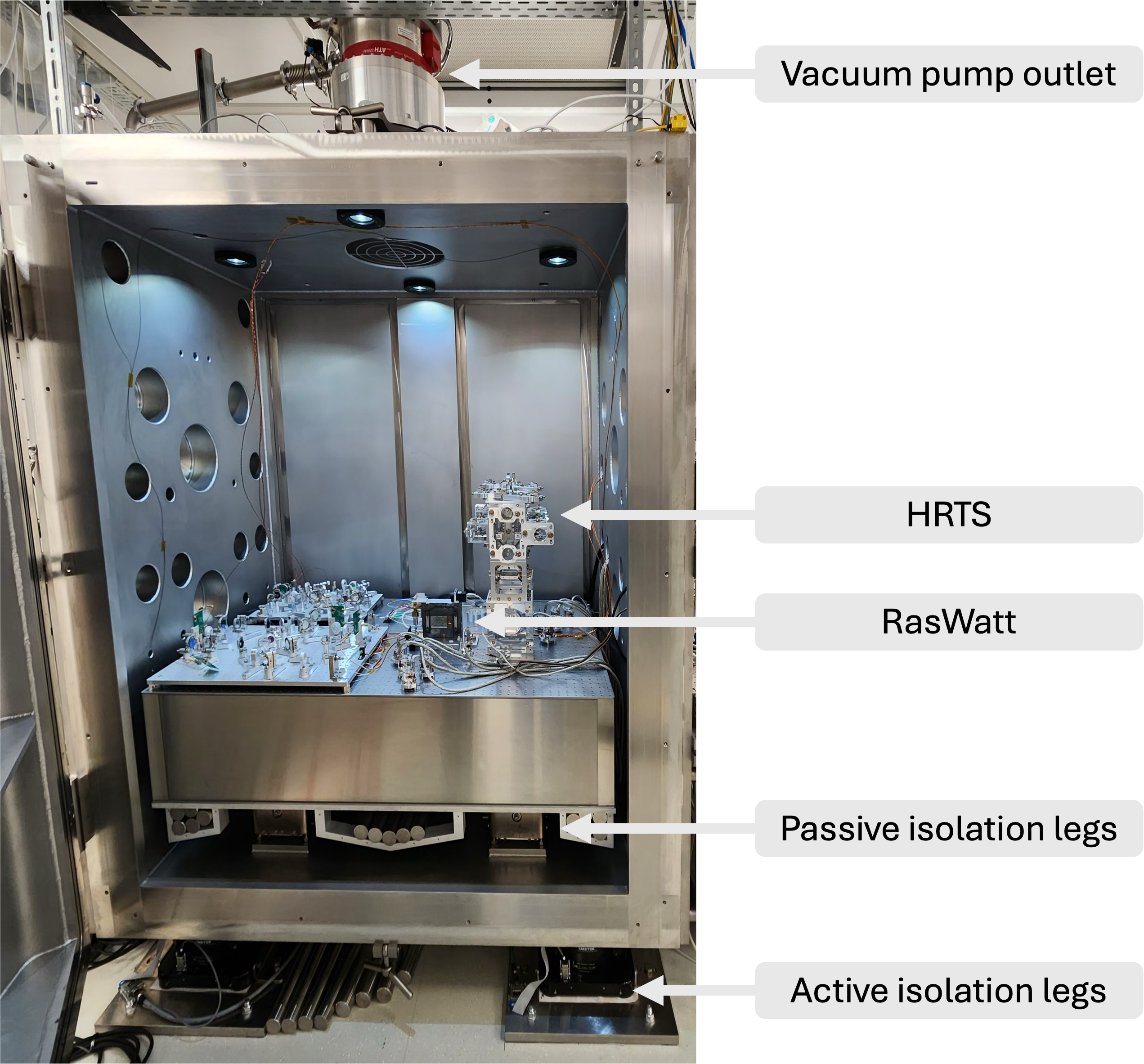}
    \caption{Experimental setup inside the VATIGrav vacuum chamber showing RasWatt mounted horizontally on the \SI{240}{\kilo\gram} optical table. The table is supported by four Minus-K CM-1 passive vibration isolators (visible as metallic legs beneath the table), while the chamber itself rests on STACIS III active isolation legs. Also visible are the HRTS (HAM Relay Triple Suspension) suspension system and the vacuum pump outlet connection on the top. This hierarchical isolation architecture combining active isolation and passive isolation creates a low-noise environment that enables separation of the optical readout floor from mechanical and seismic disturbances.}
    \label{fig:vatigrav-schematic}
\end{figure}

In this configuration, RasWatt is clamped on the passive isolation table inside the vacuum chamber, and the proof mass can be locked or released to distinguish readout floor from inertial response; resonance behavior emerges clearly in the unlocked state, while the locked state probes the intrinsic readout noise in a quiet environment. A photograph of RasWatt in the VATIGrav vacuum chamber is shown in figure~\ref{fig:vatigrav-schematic}.

The measurements were performed with active and passive isolation (denoted AI + PI) and then with only passive isolation (denoted PI).

\begin{figure}[htbp]
    \centering
    \includegraphics[width=0.8\textwidth]{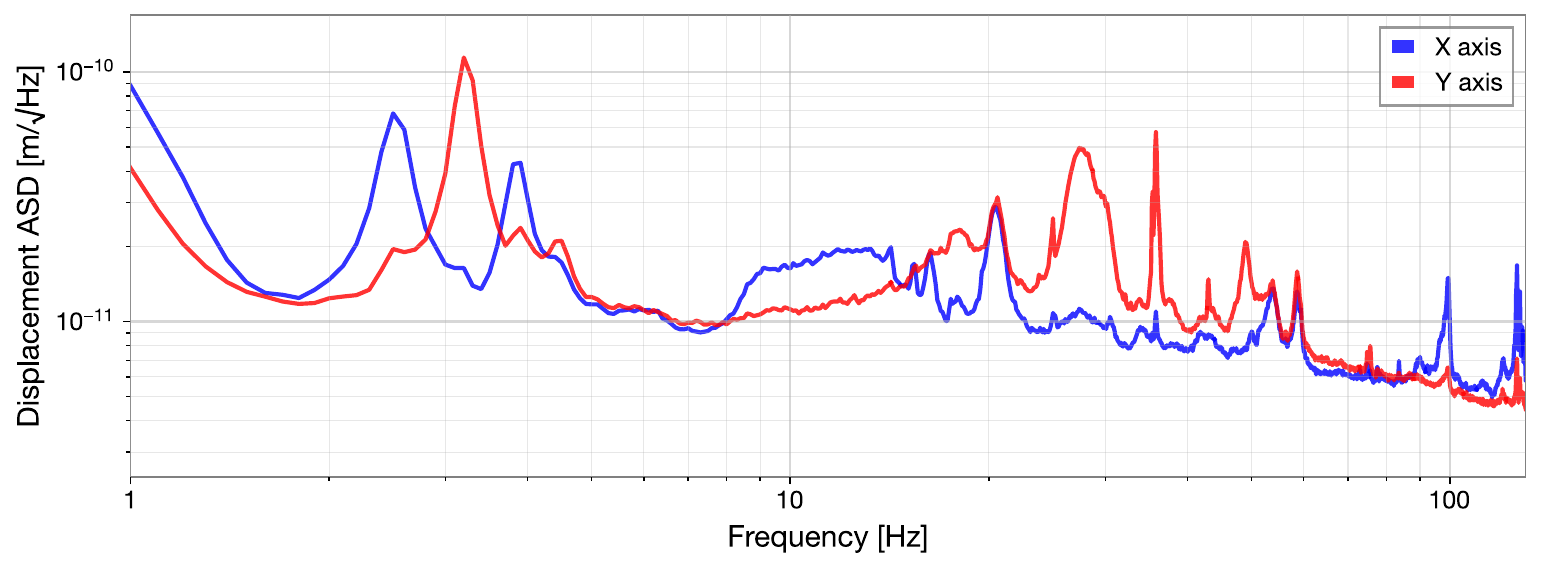}
    \caption{Locked proof-mass displacement ASD revealing the intrinsic Rasnik optical readout noise floor. Both $x$-axis and $y$-axis with active isolation on and off show comparable performance with their counterparts, approaching \SI{5}{\pico\meter\per\sqrt\hertz} at high frequencies. Active+passive isolation (AI+PI, darker) and passive-only (PI, lighter) configurations approaches \SI{5}{\pico\meter\per\sqrt\hertz}, indicating the ultimate readout performance.}.
    \label{fig:raswatt_locked_asd}
\end{figure}

The locked configuration measurements establish the fundamental noise floor of the RasWatt optical readout system. Data were acquired at \SI{274.5}{\hertz} sampling rate, with measurements performed under atmospheric pressure (before pump down) in a sealed chamber to provide temperature stability and eliminate air currents. Between \SI{5}{\hertz} and \SI{20}{\hertz}, the displacement ASD reaches below \SI{10}{\pico\meter\per\sqrt{\hertz}}, representing the intrinsic limit of the Rasnik optical readout combined with the mechanical mounting. At high frequencies, the displacement ASD approaches \SI{5}{\pico\meter\per\sqrt{\hertz}}, demonstrating the intrinsic readout sensitivity in the quiet VATIGrav environment. Distinct peaks at \SI{2.5}{\hertz}, \SI{3.5}{\hertz}, and around \SI{50}{\hertz} indicate residual mechanical resonances in the mounting structure and electronic noise sources. The comparable performance between $x$ and $y$ axes throughout the frequency range demonstrates sensor symmetry and validates the dual-axis readout capability (see figure~\ref{fig:raswatt_locked_asd}).

\begin{figure}[htbp]
    \centering
    \includegraphics[width=0.8\textwidth]{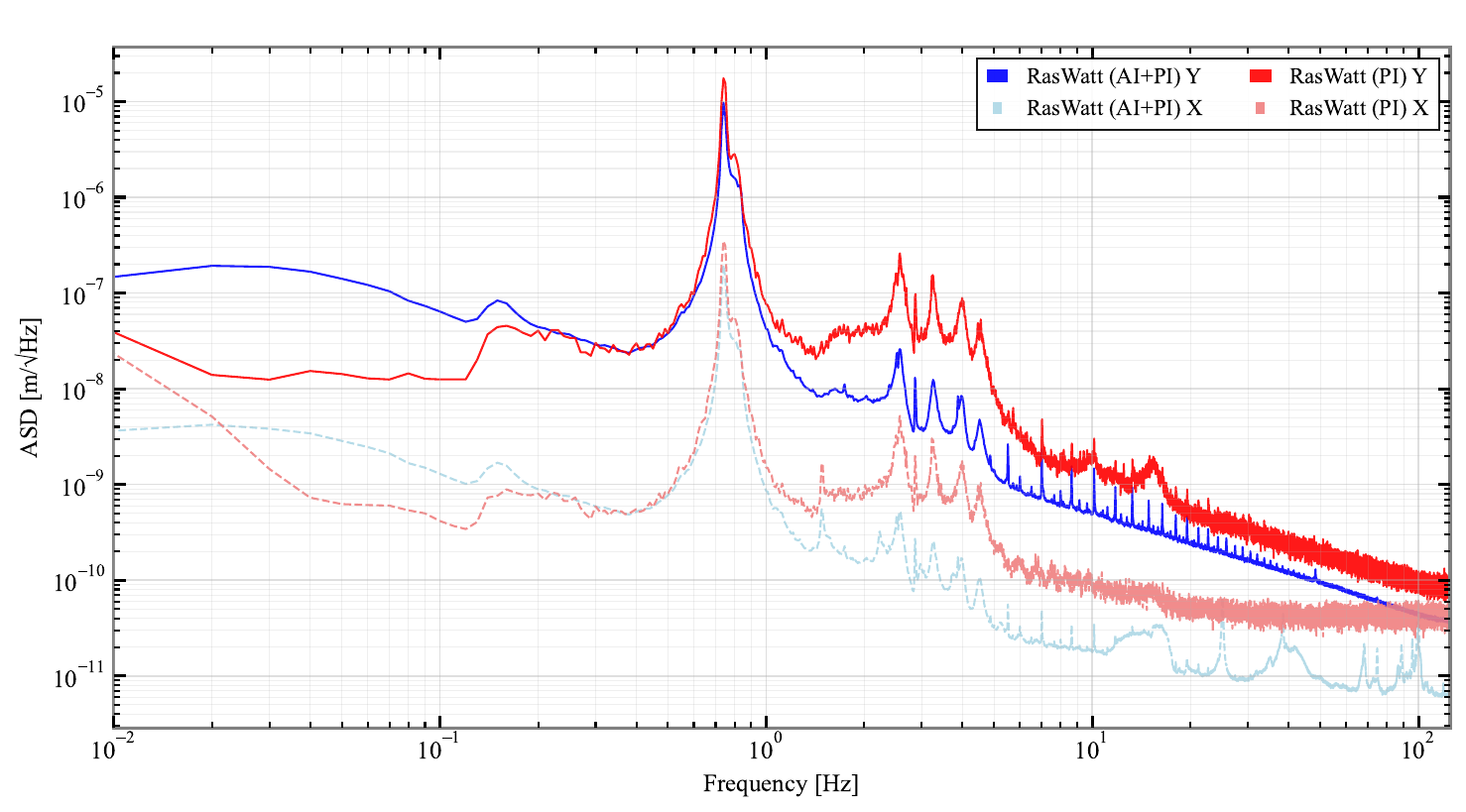}
    \caption{Unlocked proof-mass displacement ASD showing uncorrected RasWatt readout data; inertial displacement measurement. Both $x$-axis (dashed) and $y$-axis (solid) exhibit the Watt's linkage resonance at \SI{0.740}{\hertz} as a sharp peak. Active+passive isolation (AI+PI, blue/light blue) provides superior performance below \SI{5}{\hertz} compared to passive-only (PI, red/pink). Above \SI{5}{\hertz}, both configurations converge to approximately \SI{500}{\pico\meter\per\sqrt{\hertz}}. Small $y$→$x$ cross-coupling is visible in the $x$-channel resonance due to mask rotation offset during installation.}
    \label{fig:raswatt_unlocked_asd}
\end{figure}

The unlocked configuration reveals the uncorrected RasWatt displacement readout, directly measuring the proof mass motion without removal of the mechanical transfer function. The $x$ and $y$ spectra are dominated by the Watt's linkage resonance near \SI{0.740}{\hertz}, appearing as a sharp peak in the displacement measurement. This raw readout data reflects the mechanical response of the linkage to seismic input transmitted through the isolation system. The active+passive isolation configuration (AI+PI) demonstrates  lower noise in the \SI{1}{\hertz} to \SI{5}{\hertz} band compared to passive-only (PI), where passive isolator resonances (\SI{1.2}{\hertz} to \SI{2}{\hertz}) amplify residual motion. Above \SI{5}{\hertz}, both isolation configurations converge to a higher noise floor of approximately \SI{500}{\pico\meter\per\sqrt{\hertz}}. A small \(y\!\to\!x\) cross-coupling contribution is visible in the $x$-channel resonance, attributed to the mask's rotational misalignment during installation (see figure~\ref{fig:raswatt_unlocked_asd}).

\subsection{LISA Mission Verification}
\label{sec:space_applications}

The Laser Interferometer Space Antenna (LISA)~\cite{LISA2017} is one of the European Space Agency's (ESA) planned space-based gravitational wave detector, requiring picometer-level stability for its optical components. During ground-based verification testing of a specific LISA subsystem - the Quadrant Photoreceiver (QPR) which tracks inter-spacecraft laser beams, the Rasnik sensors were selected for dimensional stability measurements. Rasnik served exclusively as test equipment for characterizing this particular subsystem, not for the entire LISA spacecraft. The selection was driven by several key advantages: absolute position decoding with no cumulative drift due to absolute coding, straightforward installation and operation, and crucially, the ability to measure displacements perpendicular to the optical axis. Additionally, Rasnik's optical measurement principle exhibits minimal thermal sensitivity under controlled conditions during measurement acquisition and maintains measurement integrity when operated consistently in vacuum conditions. These characteristics make it ideal for detecting mechanical displacements of spacecraft components while remaining largely invariant to the thermal environment of the measurement system itself.

\subsubsection{Rasnik Adaptation for QPR Testing}
\label{sec:rasnik_adaptation}

\begin{figure}[htbp]
    \centering
    \includegraphics[width=0.4\textwidth]{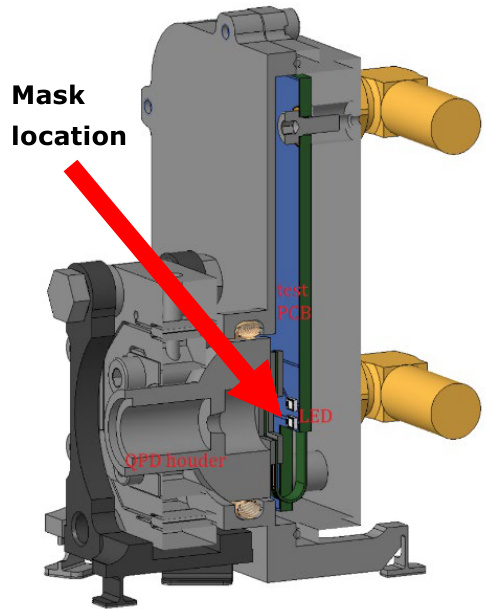}
    \caption{QPR housing cross-section (\SI{60}{\milli\meter} height) showing Rasnik mask location replacing the quadrant photodiode, connected via flex-print (green).}
    \label{fig:lisa_qpr_cross}
\end{figure}

The quadrant photodiode was replaced with a Rasnik mask bonded at the exact QPD location. figure~\ref{fig:lisa_qpr_cross} shows the QPR housing cross-section with the mask location.

The Rasnik mask consists of a coded checkerboard pattern with $9 \times 9$ element blocks enabling absolute position decoding, illuminated using a milk glass diffuser.

\begin{figure}[htbp]
    \centering
    \includegraphics[width=0.6\textwidth]{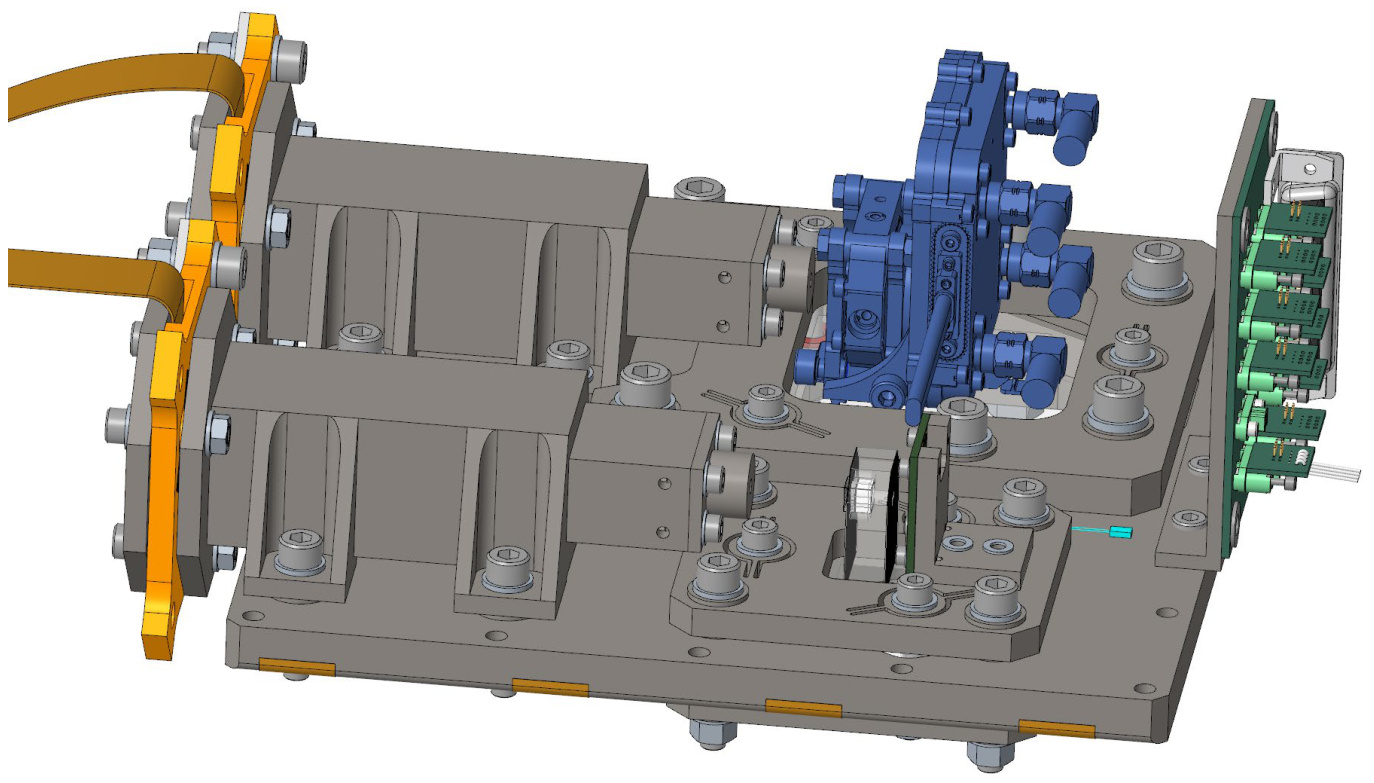}
    \caption{Dual-channel Rasnik test setup: QPR measurement (upper) and ClearCeram reference (lower) channels with Invar structure (dark grey), QPR housing (blue), cameras (orange), and PCBs (green).}
    \label{fig:lisa_setup_cad}
\end{figure}

The dual-channel Rasnik system (figure~\ref{fig:lisa_setup_cad}) monitored QPR mask displacement using blue LEDs (\SI{447.5}{\nano\meter}) with a thermally stable ClearCeram reference mask to isolate setup artifacts.

\begin{figure}[htbp]
    \centering
    \includegraphics[width=0.8\textwidth]{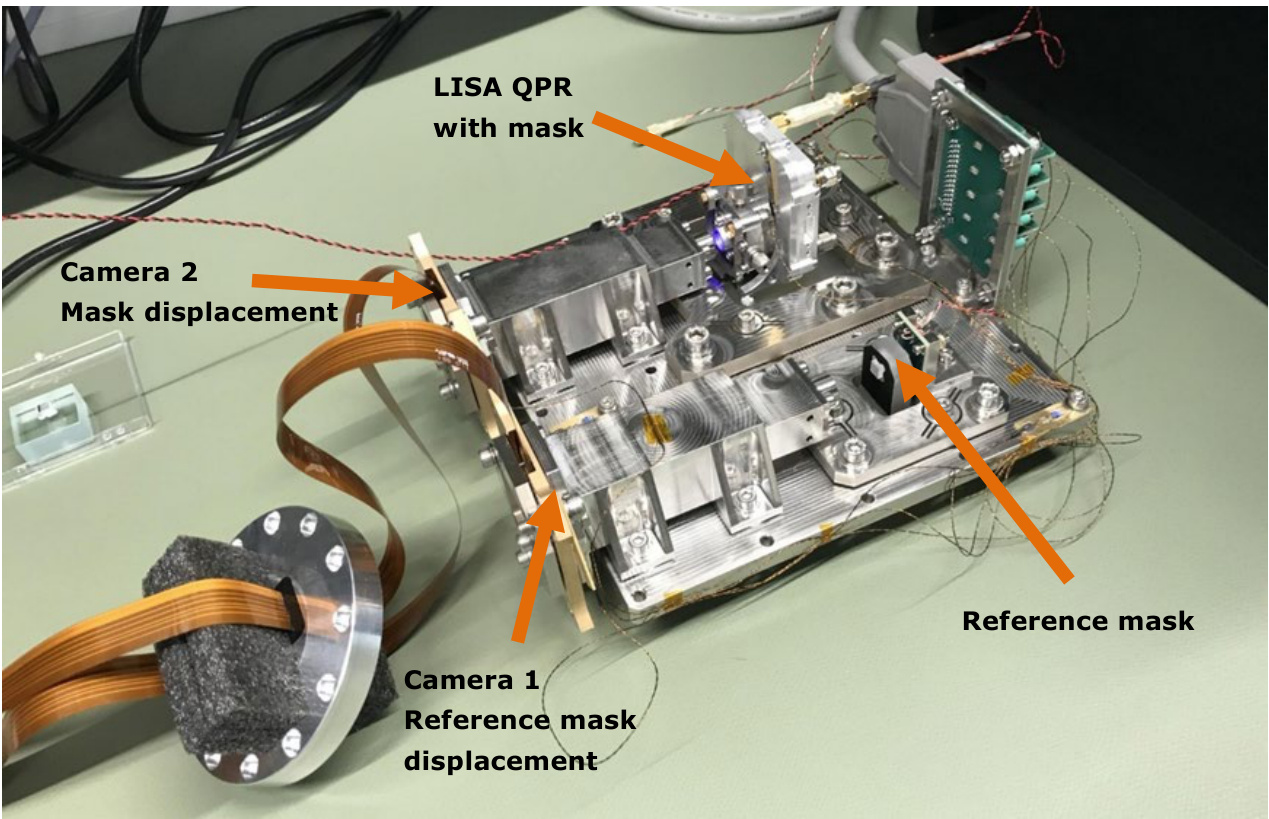}
    \caption{Photograph of the assembled Rasnik test setup with the LISA QPR housing integrated for thermal vacuum testing.}
    \label{fig:lisa_setup_photo}
\end{figure}

CMOS cameras viewed the masks through \SI{20}{\milli\meter} optical paths. The titanium-aluminum thermal center design and Invar mounting minimized thermally-induced displacements. figure~\ref{fig:lisa_setup_photo} shows the assembled configuration.

\subsubsection{Thermal Vacuum Test Results}
\label{sec:tv_results}

\begin{figure}[htbp]
    \centering
    \includegraphics[width=\textwidth]{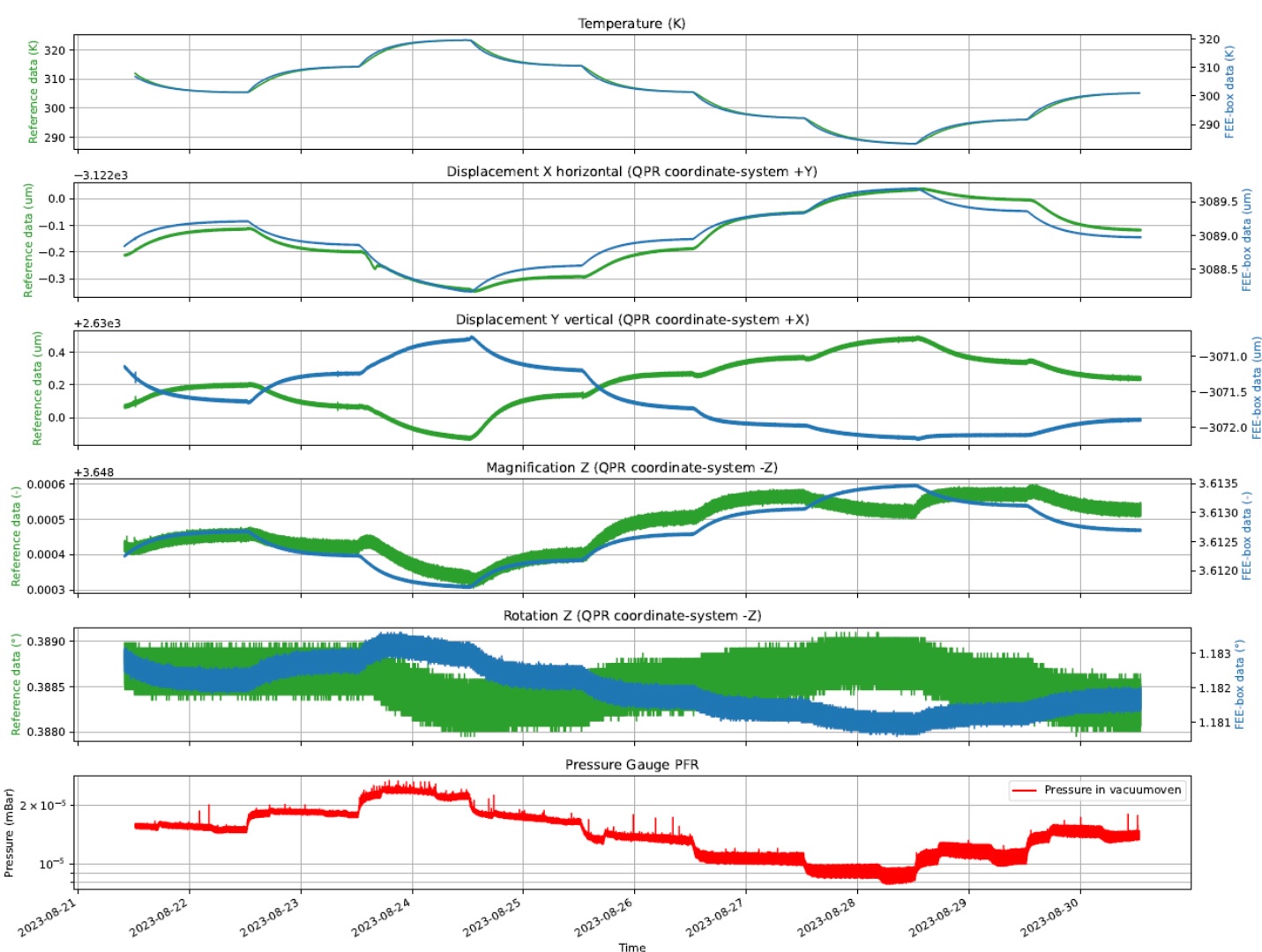}
    \caption{Time series data from the thermal vacuum test campaign. From top to bottom: temperature profile showing stepped cooling from \SI{320}{\kelvin} to \SI{290}{\kelvin} and return; horizontal displacement (X in QPR coordinates) for both QPR (green) and reference (blue) channels; vertical displacement (Y in QPR coordinates); optical magnification Z tracking thermal expansion effects; rotation angle showing minimal angular deviation; and vacuum chamber pressure maintained at approximately \SI{1e-5}{\milli\bar} throughout the test.}
    \label{fig:lisa_time_series}
\end{figure}

The QPR assembly underwent thermal cycling from \SI{0}{\degreeCelsius} to \SI{40}{\degreeCelsius} in \SI{10}{\degreeCelsius} steps under vacuum (\SI{2e-5}{\milli\bar}). RasCal measured continuous relative displacements between the QPR mask and reference. figure~\ref{fig:lisa_time_series} shows the complete time series over the multi-day campaign.

The primary findings demonstrated excellent thermomechanical stability. The horizontal displacement (Y-axis in QPD coordinates) measured \SI{-32 \pm 14}{\nano\meter\per\kelvin}, while the vertical displacement (X-axis in QPD coordinates) was \SI{55 \pm 24}{\nano\meter\per\kelvin}. Both values fall well within the LISA requirement of $\leq \SI{100}{\nano\meter\per\kelvin}$ at the mask's position during operational temperatures.

\begin{figure}[htbp]
    \centering
    \includegraphics[width=0.49\textwidth]{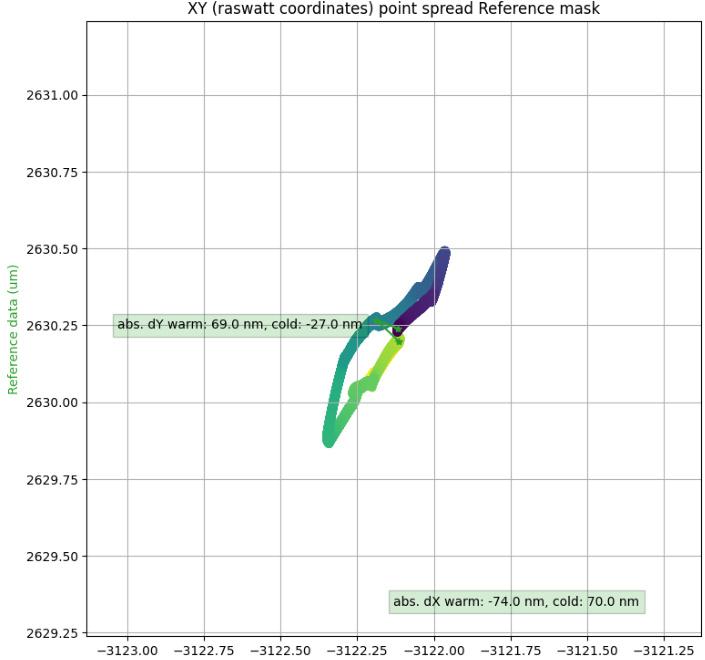}
    \includegraphics[width=0.49\textwidth]{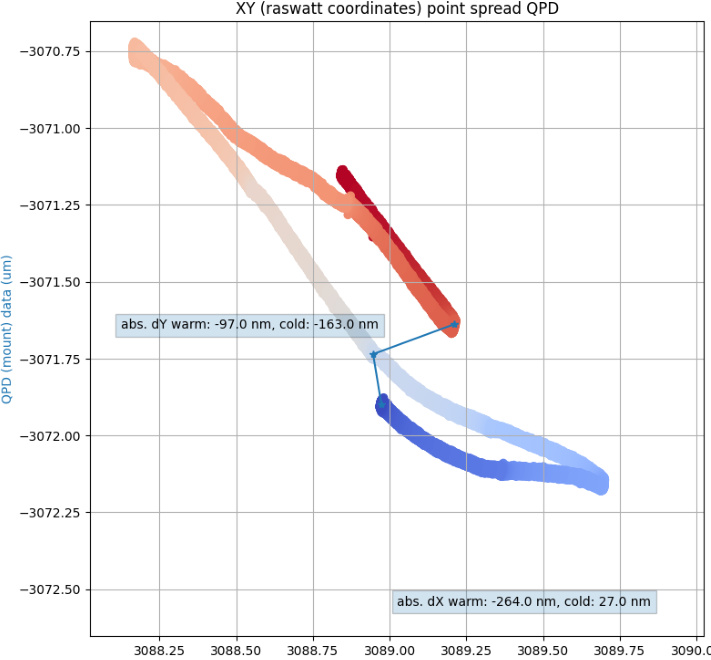}
    \caption{XY position trajectories during thermal cycling: (left) thermally stable reference mask with minimal hysteresis (\SI{74}{\nano\meter} $\times$ \SI{70}{\nano\meter}), (right) QPD position with thermal loops (max \SI{264}{\nano\meter} $\times$ \SI{163}{\nano\meter}). Color gradients show thermal cycle progression, validating mechanical stability within LISA requirements.}
    \label{fig:lisa_xy_spread}
\end{figure}

Maximum thermal hysteresis of \SI{264}{\nano\meter} was observed, meeting the \SI{1000}{\nano\meter} requirement. Figure~\ref{fig:lisa_xy_spread} shows thermal hysteresis through XY trajectories, with the reference mask exhibiting minimal hysteresis while the QPD shows characteristic thermal cycling behavior.

These measurements validated the QPR mechanical design for the LISA environment. The Rasnik system's absolute position decoding with no cumulative drift and stable operation without recalibration over multi-day campaigns established it as a valuable tool for space hardware qualification.

    \section{Conclusion}

A comprehensive control and analysis framework, RasCal, has been presented, extending the Rasnik alignment concept into a robust, near real-time metrology system. By combining absolute-value coded masks with two-dimensional spectral analysis, sub-nanometer spatial resolution without cumulative drift has been demonstrated, with simultaneous readout of multiple degrees of freedom transverse to the optical axis and sensitivity to magnification and rotation. On commodity hardware, processing rates exceeding \SI{300}{\hertz} in maximum-throughput tests and \SI{274.5}{\hertz} during live acquisition with GPU acceleration have been achieved, while CPU-only operation sustains approximately \SI{109}{\hertz}. Sub-millisecond command latency and thread-safe multi-camera operation have been realized. In quiet laboratory conditions, a displacement sensitivity of \SI{5}{\pico\meter\per\sqrt\hertz} has been obtained, indicating proximity to the limits set by photon statistics.

These capabilities have been validated in two representative contexts. In the RasWatt implementation on a Watt’s linkage, separation between readout noise and inertial response was achieved; resonance features were characterized; and velocity-dependent manifestations of residual linearity error were observed. During thermal-vacuum testing of the LISA Quadrant Photoreceiver (QPR), absolute, drift-free tracking of mask position perpendicular to the optical axis was maintained over multi-day campaigns, yielding temperature coefficients and hysteresis within the specified requirements. Together, these results support RasCal’s suitability for precision alignment tasks in gravitational-wave instrumentation and space hardware qualification.

The principal limitations are well understood. The periodic, position-dependent linearity error inherent to discrete phase extraction in the two-dimensional Fourier analysis imposes a constraint that remains above the shot-noise floor; its dynamic counterpart appears as velocity-dependent noise during motion. Both effects are substantially mitigated by introducing a small differential rotation between mask and sensor and by post-processing corrections. Reliable operation further benefits from uniform illumination, controlled magnification, and stable contrast, aspects that are addressed by the current pipeline and data-quality checks. RasCal is currently distributed on request; interested parties may contact the corresponding author for access and collaboration.

\paragraph{Outlook}
Future work will concentrate on algorithmic refinements that further suppress linearity errors, including sinc-based or maximum-likelihood peak estimation that jointly incorporates code-square constraints and rotation. Closed-loop integration with alignment controls, synchronized multi-camera operation, and sustained operation in vacuum and thermally dynamic environments are planned. Application-level developments toward differential motion monitoring between separated optical platforms and active alignment control are being explored.

\acknowledgments
We thank Nikhef, Amsterdam for providing Rasnik components (legacy hardware on which this work builds) and the University of Hamburg for access to and support with the VATIGrav facility.

\appendix
\pagebreak
\section{Abbreviations}
\begin{table}[H]
    \centering
    \begin{tabular}{c|c}
    \hline
       API  & Application Program Interface \\
       ASD & Amplitude Spectral Density \\
       CAD & Computer-Aided Design \\
       CMOS & Complementary Metal-Oxide-Semiconductor \\
       CRLB & Cramér-Rao Lower Bound \\
       CTE & Coefficient of Thermal Expansion \\
       DAQ & Data Acquisition \\
       DFT & Discrete Fourier Transform \\
       DOF & Degree of Freedom \\
       ESA & European Space Agency \\
       EWMA & Exponentially Weighted Moving Average \\
       FEE & Front-End Electronics \\
       FFT & Fast Fourier Transform \\
       FPS & Frames Per Second \\
       LED & Light Emitting Diode \\
       LISA & Laser Interferometer Space Antenna \\
       NIC & Network Interface Controller \\
       PCB & Printed Circuit Board \\
       QPD & Quadrant Photodiode \\
       QPR & Quadrant Photoreceiver \\
       RoI & Region of Interest \\
       SNR & Signal-to-Noise Ratio \\
       SPI & Suspension Point Interferometer \\
       UHV & Ultra High Vacuum \\
    \hline
    \end{tabular}
    \caption{Abbreviations used in this paper.}
    \label{tab:abbreviations}
\end{table}

\section{PC specifications}
\label{app:pc_specs}
\begin{table}[H]
    \centering
    \begin{tabular}{ll}
        \hline
        \multicolumn{2}{c}{\textbf{System Information}} \\
        \hline
        Host & uburas \\
        Kernel & 4.15.0-39-generic x86\_64 \\
        Distribution & Ubuntu 18.04.4 LTS \\
        \hline
        \multicolumn{2}{c}{\textbf{Hardware}} \\
        \hline
        System & Dell OptiPlex 3020 \\
        Motherboard & Dell 040DDP (A01) \\
        BIOS & Dell A03 (14/04/2014) \\
        \hline
        \multicolumn{2}{c}{\textbf{CPU}} \\
        \hline
        Processor & Intel Core i5-4590 (Haswell) \\
        Cores & 4 \\
        Max Clock Speed & \SI{3.7}{\giga\hertz} \\
        \hline
        \multicolumn{2}{c}{\textbf{Graphics}} \\
        \hline
        Integrated GPU & Intel Xeon E3-1200 v3/4th Gen Core \\
        Discrete GPU & NVIDIA GeForce GTX 1050 \\
        \hline
        \multicolumn{2}{c}{\textbf{Network}} \\
        \hline
        NIC 1 & Realtek RTL8111/8168/8411 (\SI{1000}{\mega\bit\per\second}) \\
        NIC 2 & Intel 82572EI Gigabit (\SI{1000}{\mega\bit\per\second}) \\
        \hline
        \multicolumn{2}{c}{\textbf{Storage}} \\
        \hline
        Drive & WDC WD5000AAKX \SI{500}{\giga\byte} \\
        \hline
        \multicolumn{2}{c}{\textbf{Memory}} \\
        \hline
        RAM & \SI{8}{\giga\byte} \\
        \hline
    \end{tabular}
    \caption{Specifications of the computer used for development and testing of RasCal.}
    \label{tab:pc_specs}
\end{table}

\bibliographystyle{JHEP}
\bibliography{refs}
    
\end{document}